\documentclass[%
reprint,
amsmath,amssymb,
aps,
prx,
floatfix,
]{revtex4-2}

\usepackage{graphicx} % Include figure files
\usepackage{dcolumn} % Align table columns on decimal point
\usepackage{bm} % bold math
\usepackage{natbib}
\usepackage{appendix}
\usepackage{color}
\begin{document}

\preprint{APS/123-QED}

\title{ Observation of Universality in Decaying Turbulence}

% Force line breaks with \\

\author{Christian K\"uchler}
 \email{christian.kuechler@ds.mpg.de}
 \affiliation{Max Planck Institute for Dynamics and Self-Organization, G\"ottingen, Germany \\
 Institute for Dynamics of Complex Systems, University of G\"ottingen, G\"ottingen, Germany}
 %Lines break automatically or can be forced with \\

\author{Gregory P. Bewley}
 \email{gpb1@cornell.edu}
 \affiliation{%
Sibley School of Mechanical and Aerospace Engineering, 
Cornell University, Ithaca, NY, USA}

\author{Eberhard Bodenschatz}
 \email{eberhard.bodenschatz@ds.mpg.de}
 \affiliation{Max Planck Institute for Dynamics and Self-Organization, G\"ottingen, Germany \\
 Physics Department,  Cornell University, Ithaca, NY, USA \\
 Institute for Dynamcis of Complex Systems, University of G\"ottingen, G\"ottingen, Germany}

\date{\today}% It is always \today, today,
             %  but any date may be explicitly specified

\begin{abstract}

A hallmark of fluid turbulence theory is the universal power law scaling of the velocity difference statistics between two points in space in the inertial range between the large energy injection scale and the small energy dissipation scale. Even at the highest Reynolds numbers available, laboratory and natural flows such universal power laws have not been convincingly demonstrated. Here we show for the decaying active grid turbulence of the Max Planck Variable Density Turbulence Tunnel \cite{Bodenschatz2014,Griffin2019} that the velocity difference statistics at high Reynolds numbers do not exhibit a power law, but have a universal functional form independent of the Reynolds number. We separate this functional form from the power law exponent and discuss potential consequences for turbulence modelling. 

\end{abstract}

\pacs{Valid PACS appear here}% PACS, the Physics and Astronomy
                             % Classification Scheme. 
%\keywords{Suggested keywords}%Use showkeys class option if keyword
                              %display desired
\maketitle

%\tableofcontents

%--universality: what is observed in the atmosphere is also observed in a mixer. 

%--universality with respect to changes in the Reynolds number. 

%--universality in the sense of small-scale structure at high Reynolds numbers. 

%--examine specifically the velocity differences as a proxy... 

%--observe X and X

%--employ a physically motivated but incorrect model to estimate exponents

%"The essential feature is that it includes an influence of the large scales towards inertial range to explain some of the anomalous behaviour we observe. "

%Turbulence in a three-dimensional, incompressible fluid 
%an be described by a flux of kinetic energy 
%rom large energy injection length scales $L$ to small viscous scales $\eta$, where 
%internal friction dissipates this kinetic energy into heat. 
%at small scales scales described by the viscous length scale $\eta$. 
%%statistics of the turbulent velocity fluctuations include the moments of velocity differences. 
Turbulence in a three-dimensional incompressible fluid can be described by a flow of kinetic energy from large energy injection length scales $L$ to small viscous scales $\eta$, where internal friction dissipates this kinetic energy into heat.  For intermediate scales, i.e., in the inertial range, the statistics of turbulent velocity fluctuations are described  by the moments of two-point velocity increments \cite{Kolmogorov1941}. 
The  $n$-th order moments of velocity increments are called structure functions, $S_n(r) = \langle (u(x+r)-u(x))^n \rangle$. 
The separation between large and small scales, or the size of the  inertial range, 
goes hand in hand with the magnitude of 
the main parameter capturing the intensity of a turbulent flow, which is the Taylor-scale Reynolds number
$R_\lambda = u_{RMS}\lambda/\nu$. 
$u_{RMS}$ is the root-mean-squared velocity fluctuation, 
$\nu$ is the kinematic viscosity of the fluid, 
and $\lambda$ is the length scale defined in \citet{Taylor1935}, where $L \gg \lambda \gg \eta$. 
%For large $R_\lambda$, it is expected that the physical processes dominant at any given scale (and thus the scale-by-scale statistics) are universal, i.e. independent of the particularities of the what drives the turbulence. 
%An often used quantity to test this are the $n$-th order moments of the velocity increment called ``structure functions'' $S_n(r) = \langle (u(x+r)-u(x))^n \rangle$. 

\begin{figure*}[ht]
		\includegraphics[width =\textwidth]{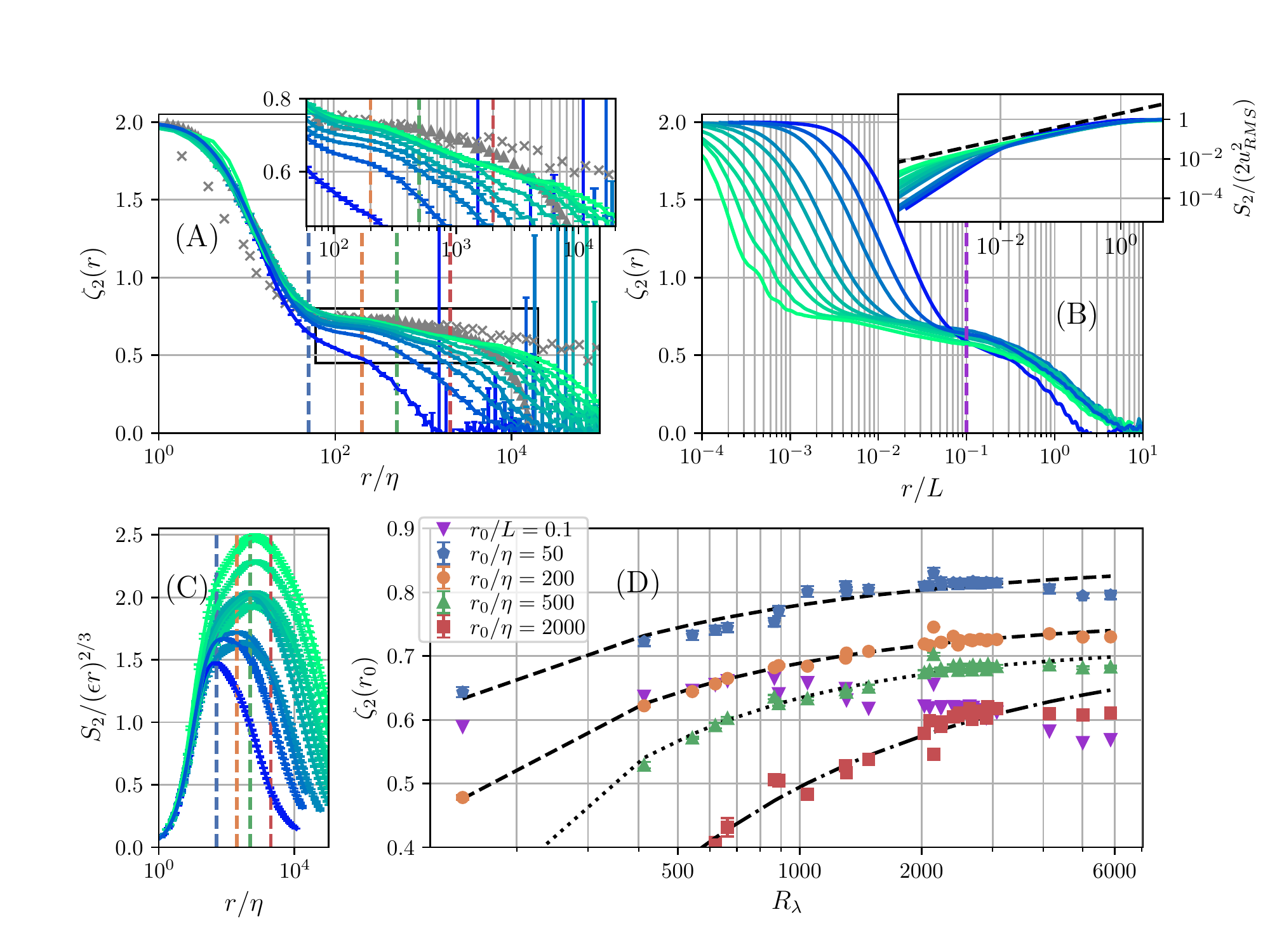}
        \caption{(A): $\zeta_2(r)$ for $R_\lambda$ = 150, 410, 660, 890, 1480, 2030, 2680, 3070, 4140 and 5860. 
        The curves collapse approximately onto a universal form for $R_{\lambda}>2000$ at scales extending up to 1000$\eta$ ($\approx 0.1 L$) as seen in the inset. 
        This form extends from the smallest scales up to $0.1L$ and 
        is different from a constant, which indicates that power law scaling is masked in these data. 
        In contrast, the curves at $R_{\lambda}<2000$ change shape significantly with $R_{\lambda}$. 
        Grey triangles: direct numerical simulations at $R_\lambda=2250$ \cite{Ishihara2020}. Grey crosses: atmospheric measurements \cite{Tsuji2004}. 
        \textit{Inset:} Zoom on the inertial range of the same curves. 
        %At the largest $R_\lambda$ a wave-like fine structure can be seen as in \citet{Sinhuber2017}. 
        Dashed lines: $r_0/\eta$ for the curves in (D). 
        (B): Same as (A), but normalised by $L$. At the largest scales the curves 
        follow a similar shape 
        %approach an approximately universal form 
        from the largest scales down to $0.2L$. 
        Dashed line: $r_0/L$ for the curve in (D). \textit{Inset}: $S_2$ normalised by its large scale expectation value of $u_{RMS}^2$. Dashed line: $\sim r^{2/3}$. The transition from the approximate inertial range to the large-scale dominated range is at $r/L>0.1$. 
        (C): Structure functions $S_2$ compensated by the scale-invariant prediction, 
        $(\varepsilon r)^{2/3}$. 
        (D): $\zeta_2(r)$ evaluated at fixed $r_0/\eta$ given by the dashed lines in (A), 
        and fixed $r_0/\eta$ given by the curve in (B). Error bars smaller than symbols in most cases. 
        To the extent that the curves approach constants, 
        these constants depend on $r_0$. 
        Therefore, no single scaling exponent $\bar{\zeta_2}$ can be isolated. 
        Dashed lines are fits of $\alpha_1 - \alpha_2 R_{\lambda}^{\beta}$ to the data. Note that the inertial range as we define it here extents from $\approx 100 \eta$ to $\approx 0.1L$.
        Where shown, vertical errorbars indicate a 95\%-CI from 30- or 100-second sections of the hours-long datasets and a 4\%-error on the mean velocity in $r=U\Delta t$. 
        %The yellow inverted triangles correspond to the dashed line in (B) and show the scaling exponent at a fixed scale relative to $L$ instead of $\eta$. 
        %Black points above this curve are within the inertial range (except for the case $r_0/\eta=10$).
        }
		\label{LocalSlopeCollapseEta}
\end{figure*}

A fundamental question is to what extent turbulent self-organisation leads to universal  statistics such as $S_n(r)$. In this context universality is understood as the collapse of statistics for flows of very different origin independent of Reynolds number upon proper normalisation. For example, in the inertial range  the statistics of velocity increments should be the same for jets, wind tunnels, mixing flows, the atmospheric boundary layer or any other in-compressible flows at sufficiently large Reynolds number. This conjecture is closely related to Kolmogorov's seminal 1941 work \cite{Kolmogorov1941} where he posited for statistically isotropic, homogeneous turbulence  that the flow-specific energy injection mechanisms impact the statistics only at large scales $\sim L$, but universal self-organisation prevails at scales $r \ll L$ down to the dissipation scale $\eta$. In this 
%For intermediate scales $L \gg r \gg \eta$, where the fluid inertia dominates over viscous forces, \citet{Kolmogorov1941} described 
%this universality in terms of the structure functions $S_n(r)$ defined above. 
%in statistically homogeneous and locally isotropic turbulence at large Reynolds numbers. 
inertial range, 
%velocity difference statistics are parameterised by 
the mean power per unit mass, $\varepsilon$, 
describes the energy transfer from energy injection scales to dissipative scales. 
Dimensional analysis then yields universal scaling laws for the inertial range structure functions
\begin{align}
    S_n(r) =& \; C_n (\varepsilon r)^{\zeta_n} \label{eq:PowerLaw}\\
    \zeta_{n,K41}=& \; n/3 
    \label{eq:K41}, 
\end{align}
where $C_n$ are universal constants in \citet{Kolmogorov1941}. % and only $C_3$ = $-4/5^{ths}$ is known. 
% they are universal according to K41 -- I am not sure what you meant by "presumably." this para introduced K41. - Agreed. CK

\begin{figure}
    \centering
    \includegraphics[width=\columnwidth]{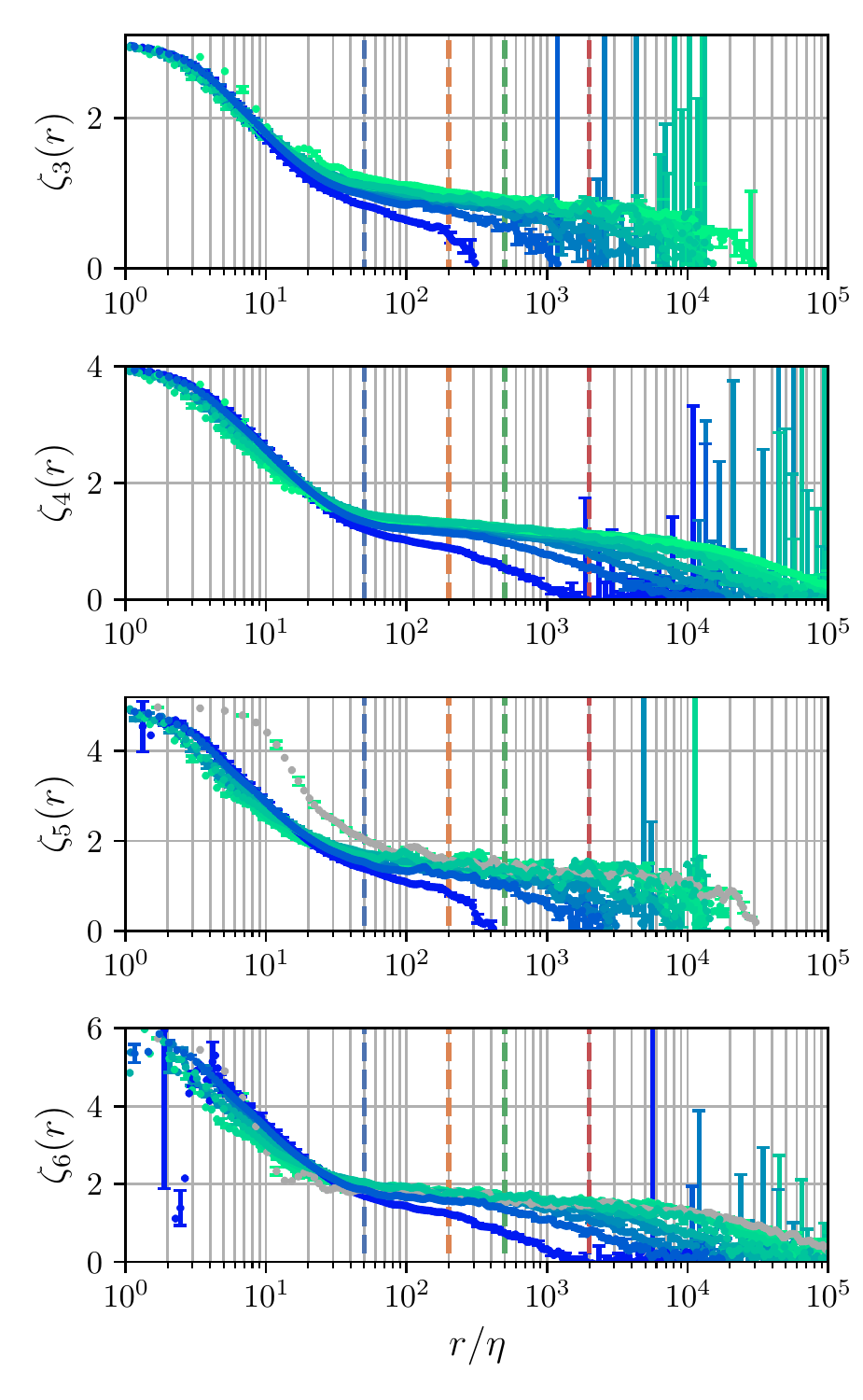}
    \caption{Same as Fig.~\ref{LocalSlopeCollapseEta} (A) 
    but for orders $3<n\leq 6$, 
    showing that the general trends observed at the second order are preserved at higher orders. 
    $\zeta_5$ was smoothed using cubic splines, and those data do not converge 
    as well at the highest $R_\lambda=5890$ in gray are likely influenced by a non-constant frequency response at small scales. 
    % is influenced by instrumentation limitations. 
    }
    \label{fig:ZetanvsR}
\end{figure}

\begin{figure}
    \centering
    \includegraphics[width=\columnwidth]{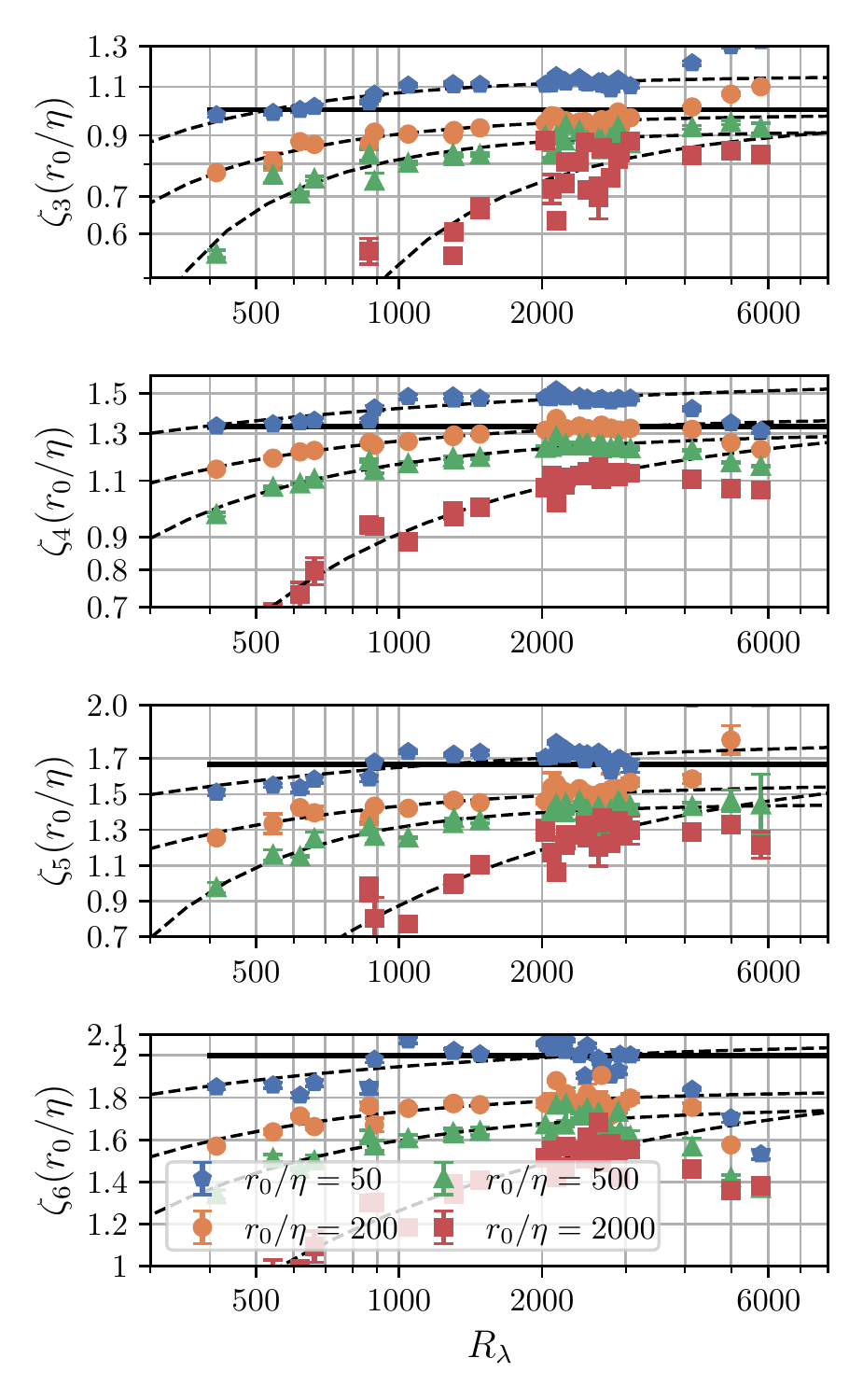}
    \caption{Same as Fig.~\ref{LocalSlopeCollapseEta} (D) 
    but for $3<n \leq 6$, 
    showing as in Fig.~\ref{fig:ZetanvsR} that the trends observed at 
    the second order are visible also at higher orders. 
    The solid black lines show the result of Kolmogorov's \cite{Kolmogorov1941} dimensional analysis $\zeta_n=n/3$, which lies above the data for values of $r_0$ in the inertial range. Dashed lines are fits of $\alpha_1 - \alpha_2 R_{\lambda}^{\beta}$ to the data excluding the largest three $R_\lambda$. 
    }
    \label{fig:ZetaR0vsRlam}
\end{figure}

The K41 scaling laws are still widely used approximations \cite{Meyers2004}, even though we know that the intermittent spatial distribution of dissipation demands corrections \cite{Batchelor1947,G.K.Batchelor1949,Kolmogorov1962}. 
% "first order approximation" not clear: S2 and the spectrum for instance are second order quantities. "these" at the beginning of a para is unclear. I would cite batchelor here or the empirical work that inspired the theory. - Agreed. CK. 
While much effort has been invested in modelling intermittency corrections to the scaling exponents $\zeta_n$ \cite{Kolmogorov1962,Frisch1978,Benzi1984,Sreenivasan1986,Meneveau1987,andrewsStatisticalTheoryDistribution1989,She1994,Dubrulle1994}, the approach towards universal, $R_\lambda$-independent scaling laws at $R_\lambda \rightarrow \infty$ is rarely questioned. 
An exception is the work by \citet{Barenblatt1995}, where a continued but potentially universal $R_\lambda$-dependence is assumed. 

The scaling laws (\ref{eq:PowerLaw}) have been derived under the idealised assumptions of a statistically stationary, homogeneous and isotropic flow in the limit of $\nu \rightarrow 0$. %Homogeneity and isotropy are severely violated in most natural flows and can only be  poorly approximated in laboratory experiments. 
Direct numerical simulations (DNS) permit the study of non-decaying isotropic turbulence as the turbulence is forced continuously in the bulk.  In both DNS and experiments, building controlled high Reynolds number turbulent remains a practical challenge. Adequately large Reynolds numbers were  available up to recently only in natural atmospheric flows, which are inhomogeneous and non-stationary, or turbulence in (super)fluid helium \cite{Praskovsky1994,Sreenivasan1998,Kahalerras1998,Tsuji2004a}, where non-intrusive measurements are extremely challenging due to the small viscous length scales \cite{White2002,Pietropinto2003,Bewley2009,Salort2012,Rousset2014}. 

\begin{figure}
  \includegraphics[width=\columnwidth]{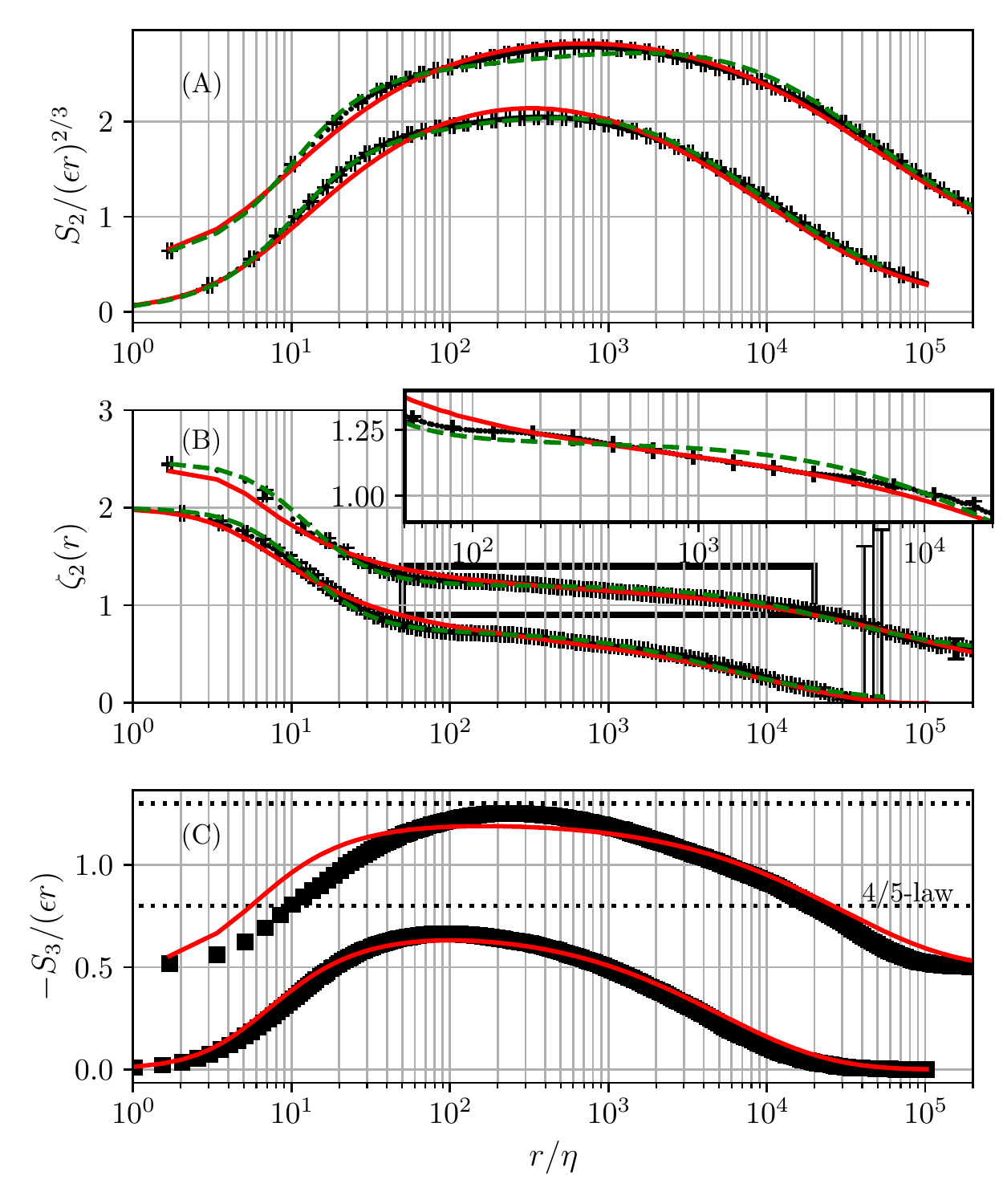}
  \caption{Comparisons of models (red, green) with experimental data (black) 
  for two Reynolds numbers, $R_\lambda = 1300$ (lower curves) and $R_\lambda = 4140$ (upper curves, offset by 0.5 for clarity). 
  Batchelor's formula is shown as green dashed lines, 
  and the model in \citet{Yang2018} as red solid lines. 
  (A) Fits of models to $S_2$ compensated by Kolmogorov's prediction eq.~(\ref{eq:K41}). 
  At large $R_\lambda$, the Batchelor formula assumes a more pronounced 
  inertial range plateau than the data. 
  (B) For power laws, 
  the region of interest is the inertial range ($100 < r/\eta < 10000$ in the high-$R_{\lambda}$ case) as highlighted by the inset, and where the red curves follow the data more closely. 
  For $r/\eta < 100$ the Batchelor interpolation is superior, 
  and the \citet{Yang2018} model performs poorly as expected. 
  At scales larger than those in the inertial range, both fits have a similar quality. 
  (C) Fits of models to $-S_3$ compensated by Kolmogrov's prediction eq.~(\ref{eq:K41}). 
  The high-$R_\lambda$ asymptote (4/5) in the inertial range is indicated by horizontal dotted lines. 
  %Structure functions approach their asymptotic values at lower $R_\lambda$ at odd orders. 
  %We observe that the model asymptotically approaches an inertial range plateau, whereas the experimental data does not. 
  }
  \label{fig:Fits}
\end{figure} 

%As first principle calculations are not yet able to elucidate the expected universality only experiments and numerical simulations can provide deeper insights. 
A large body of experimental and numerical data is available at lower $R_\lambda$. 
At these $R_\lambda$ and at low orders $n$, $S_n \sim r^{n/3}$ 
only approximates the existing data \cite[e.g.][]{Saddoughi1994,Mydlarski1996,Antonia2019}. 
%(see Refs.~\cite{Saddoughi1994,Mydlarski1996} and \cite{Antonia2019} for representative examples). 
This is because viscosity influences relatively large scales compared with the dissipation scale through the so called bottleneck \cite{Sinhuber2017,Kuchler2019} shadowing the inertial range scaling in both experiments and numerical simulations \cite{Fukayama2000,Gotoh2002,CHEN2005,Tang2017,Yeung2018}. 

In all experimental flows known to us the turbulence is generated locally in  space and is decaying away from the source of turbulence. This is true for wind tunnels, wakes, jets, counter-rotating disks, vibrating grids, etc. These effects are known to adversely affect the buildup of power law scaling in the inertial range \cite{Danaila2002,Antonia2015,Antonia2019}. The effects of decay and anisotropic energy injection are typically stronger than those of the scale-local forcing in numerical simulations \cite{Fukayama2000,Danaila2002,Bos2012}. 

The effect of viscosity and the time-dependence of the flow on the velocity increment statistics can be assessed by statistically averaging Navier-Stokes equations. Assuming isotropy and homogeneity, but allowing for a term $P$ describing the time-dependence of a continuous forcing or decay, the Karman-Howarth-equation links structure functions of orders 2 and 3: \cite{deKarman1938,Antonia2003,Danaila1999,Antonia2006}
\begin{equation}
    -S_3(r) = \frac{4}{5}\varepsilon r - 6 \nu \frac{\partial S_2}{\partial r} - P.
    \label{KHE}
\end{equation}

 $P$ depends strongly on external conditions. 
The form of the statistics is typically written as 
\begin{equation}
    S_n=C_n~(\varepsilon r)^{n/3} \left(\frac{r}{L}\right)^{\mu_n} F_n(R_\lambda, r/\eta). 
    \label{eq:FRNE}
\end{equation}
Using closure models for the statistical evolution equations \eqref{KHE} \cite{Yang2018,Bos2012,Thiesset2013}, empirical parameterisations for $F_n$ \cite{Batchelor1951,Lohse1995,Kurien2000,Dhruva2000,Meyers2008}, or physically motivated derivations of the large-scale terms \cite{Antonia2003,Thiesset2013,Tang2017,Yang2018}, functional forms for $F_n$ can be found to describe data. Extensive theoretical \cite{Danaila2002,Danaila1999,Antonia2003} and experimental \cite{Antonia2019,Antonia1982,Tang2019,Antonia2006} efforts have been invested into the description of decay phenomena in this context. The results indicate that a dependence on $R_\lambda$ may not vanish before $\mathcal{O}(R_\lambda) = 10^4$ in decaying turbulence behind a passive grid. 
%** Dhruva and Sreeni spent some time developing "physical motivation" for their models. i am not sure they would like their work to be called ad-hoc. GB. I refined the wording. Their interpolation is different in spirit though, I think. 

%\begin{figure*}
%    \centering
%    \includegraphics[width=\textwidth]{Fig2.pdf}
%    \caption{Same as Fig.~\ref{fig:Fig1}, but for orders $n=3$. As %for $n=2$, universality can only be identified in $\zeta_3(r)$. %It is clear in (C) that the data does not approach a power law, %whereas a model for confined decaying turbulence predicts such a %power law.  
%     }
%    \label{fig:Fig2}
%\end{figure*}

In this article we show how velocity increment statistics approach a inertial range
that is independent of the Reynolds number above $R_\lambda \approx 2000$ up to the experimental limit of $R_\lambda \approx 6000$. From this we conclude 
that $F_n(R_\lambda,r/\eta)$ is a non-trivial,  $R_\lambda$-independent and universal function at high Reynolds numbers in decaying turbulent flows. 

%%%%%%%%%%%%%%%%%%
%%Main Body%%%%%%%
%%%%%%%%%%%%%%%%%%
%\section{Main}
We conducted experiments in the Max Planck Variable Density Turbulence tunnel   which has a volume of 88m\textsuperscript{3} and is pressurized with sulphur hexaflouride  (SF\textsubscript{6}) at pressures between 1 and 15 bar, where an approximately homogeneous central region exists within the tunnel \cite{Bodenschatz2014}. 
The turbulence was generated by an active grid, which dynamically blocks parts of the tunnel cross-section at variable length- and time scales  \cite{Griffin2019,Kuchler2019}. We have compared our data to that from a traditional passive grid in the same facility and they agree very well. 
We recorded time series of the streamwise velocity component using subminiature hot wires (Nanoscale Thermal Anemometry Probes, NSTAPs) \cite{Vallikivi2011} and conventional hot wires. The wire lengths were $\lesssim 4 \eta$. 

In eq.~(\ref{eq:FRNE}) % this equation comes later: out of order. 
we observe that the prefactors $C_n$ as well as the $r$-dependent remainder may depend on $R_\lambda$. To separate the scaling of the $n$-th order structure function $S_n$ from the constants $C_n$ we consider first the local power-law exponents 
\begin{equation}
    \zeta_n(r) = \frac{d \log(S_n)}{d \log(r)} = \frac{n}{3}+\mu_n+\frac{d \log(F)}{d\log (r)}
    \label{eq:DefLocalSlope}
\end{equation}
Fig.~\ref{LocalSlopeCollapseEta} exemplifies our results at the second order for selected $R_\lambda$. 
In panels (A) and (B) we show the local power-law exponent $\zeta_2(r)$ with the scale $r$ normalised by the viscous length scale $\eta$ and the energy injection scale $L$, respectively. 
A power law prevails 
when $\zeta_n(r)$ assumes a constant value $\zeta_n=n/3+\mu_n$, 
which is the scaling exponent. 
We find that for small $r \approx \eta$, $\zeta_2 \sim 2$ ($S_2 \sim r^2$), 
as expected from continuity. 
Around $r \approx 100\eta$, $\zeta_2(r)$ flattens as expected for the inertial range. 
The width of this approximate plateau increases with $R_\lambda$, with a tilt evident even at the largest $R_\lambda$. This shape appears not to change  starting around $R_\lambda \approx 2000$ and above. The tilt is also observed in recent DNS at $R_\lambda=2250$ \cite{Ishihara2020} and atmospheric measurements at much larger $R_\lambda\approx 17000$\cite{Tsuji2004}, but is slightly less pronounced compared to our data. Due to these properties we define the approximate plateau in $\zeta_2$ as the inertial range for the remainder of this article. 
At yet larger scales, $\zeta_2(r)$ approaches zero, its large-scale limiting value for even $n$. 
Panel (C) of Fig.~\ref{LocalSlopeCollapseEta} shows the corresponding structure functions $S_2(r)$ compensated by the Kolmogorov prediction eq.~(\ref{eq:K41}). 
No clear plateau can be observed even at the largest $R_\lambda$ indicating the absence of plain scaling. 
To better illustrate the $R_\lambda$-dependence of the local power law exponent $\zeta_2(r)$ we plot its value at specific scales $r_0$ within the inertial range as functions of $R_\lambda$ in Fig.~\ref{LocalSlopeCollapseEta} (D). 
Overall, $\zeta_2(r_0/\eta)$ reaches a constant for $R_\lambda>2000$ and any fixed $r_0$ in the inertial range. Therefore, the shape of $(r/L)^{\mu_n} F_n(r/\eta)$ in the inertial range becomes independent of $R_\lambda$ for $R_\lambda>2000$. 
However, the particular asymptotic values of $\zeta_n(r_0/\eta)$ found at each specific scale $r_0/\eta$ in the inertial range differ by up to 0.2 -- far more than typical intermittency corrections. 

The above observations apply also at higher orders, which are shown in Figs.~\ref{fig:ZetanvsR} and \ref{fig:ZetaR0vsRlam}. At the largest $R_\lambda$ and smallest scales observed, the data are likely influenced by insufficient instrument frequency responses. This is particularly important at higher orders. 

So far, we have found that we cannot infer a single inertial range exponent, and thus cannot disentangle $\mu_n$ from $F$ given a single structure function. We therefore turn to a model of decaying turbulence in a finite domain to aid us in separating $F$ and $\mu_n$. We compare the results from this analysis to an established empirical method to extract $\mu_n$.

%In the present experiments turbulence is excited by the active grid and the turbulent kinetic energy decays downstream (see Fig. \ref{fig:DecayPlot} in the Appendix). 
In freely decaying turbulence, the energy injection scale $L$ grows over time \cite{Saffman1967,Sinhuber2015}. 
In the VDTT, however, the growth of $L$ is limited by the dimensions of the wind tunnel's cross section.  Decaying turbulence in a confined domain was recently modeled by Yang, Pumir and Xu \cite{Yang2018}.  The authors derive the functional forms for the viscous and large-scale cutoffs of inertial range power laws from a closure theory and self-similar decay laws (see Methods for details). 
In the model the effective scaling exponent of the second order structure function ($n/3+\mu_{2F}$) is one parameter, while the other describes the decay and is related to the normalised rate of dissipation $C_{\varepsilon}=\varepsilon L/u^3$. The model can thus be used to separate the inertial-range scaling from large-scale effects in the present experiments. An alternative is the ad-hoc formula in Refs \cite{Batchelor1951,Dhruva2000}, which provides smooth transitions between the different scaling regimes ($r^n$,$r^{\zeta_n}$,$r^0$), but no physical justification. 

In Fig.~\ref{fig:Fits} we show the two models in red (\cite{Yang2018}) and green (\cite{Dhruva2000}) with parameters fitted to the experimental data. 
The fits indicate that the model for decaying turbulence in a confined domain \cite{Yang2018} is a better approximation at higher Reynolds numbers than the Batchelor interpolation formulation\cite{Batchelor1951,Dhruva2000}, whereas the Batchelor formula describes the data better at lower $R_\lambda$. 
Both models asymptotically approach power laws in the inertial range at very large $R_\lambda$. 
At second order the model in \citet{Yang2018} better predicts the sustained influence of turbulence decay down to relatively small scales and is close to the data in the inertial range. At third order, the model performs well only at the smaller $R_\lambda$ chosen. At large $R_\lambda$, the model is already close to its asymptotic state of $S_3/(\varepsilon r)=\mathrm{const.}$ by construction in the inertial range. This asymptotic state differs qualitatively from the behaviour we observe, which explains the differences between the model and our data.

We interpret the model in \citet{Yang2018} as a physical model for $(r/\eta)^{\mu_n} F_2(R_\lambda,r/\eta)$ and extract the intermittency correction $\mu_n$ from the data.  

\begin{figure}
  %\begin{center}
  \includegraphics[width=1\columnwidth]{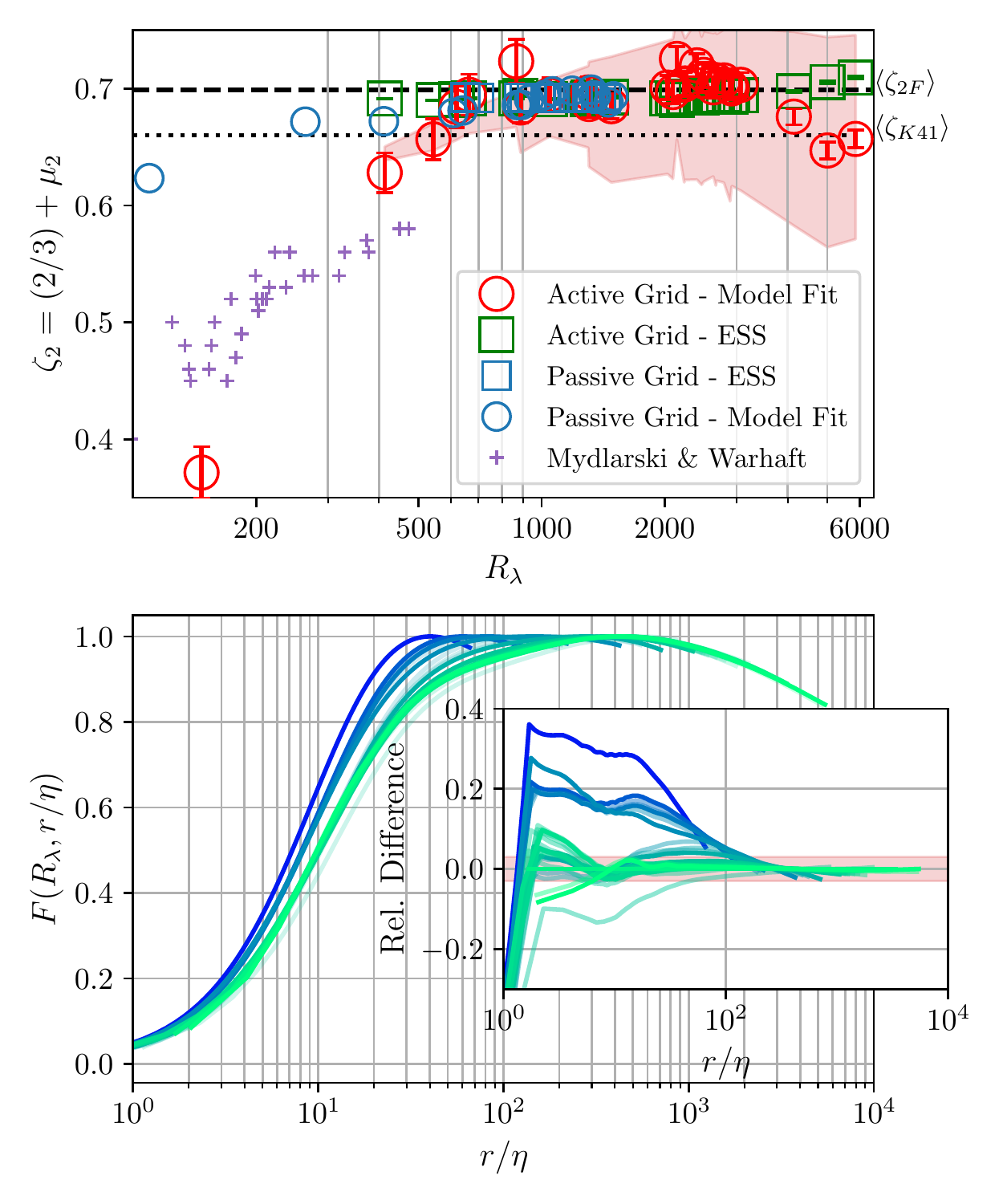}
  \caption{\textit{Upper Plot:} 
  The second-order scaling exponent $\zeta_2$ measured in different ways 
  and in different laboratory experiments.  
  Circles show $\zeta_2$ found by fitting eq.(\ref{eqn:YangModel}) to data from active and passive grid experiments. 
  Squares show extended self-similarity (ESS) exponents, $S_2(|S_3|)$, for the same datasets. 
  According to the model fits, 
  $\zeta_2$ approaches a constant $\langle\zeta_{2F}\rangle = 0.698\pm0.011$ (dashed) 
  larger than Kolmogorov's prediction (dotted) \cite{Kolmogorov1941}. 
  We attribute the slight downward trend in the last two data points to probe effects and the anisotropic forcing that we used to reach these high $R_\lambda$. 
  For comparison we include data from \citet{Mydlarski1996}. For $R_\lambda<300$, no ESS exponent could be measured due to an insufficient inertial range. 
  The shaded region corresponds to the range of values that the local slope $\zeta_2(r)$ takes within $100\eta < r < 0.1L$. 
  \textit{Lower Plot:} 
  Approach of $F_2$ measured by eq.~(\ref{eq:DefF}) towards an $R_\lambda$-universal shape. 
  Starting around $R_\lambda\approx 1500$ the curves collapse for $r<0.1L$. 
  %, show universality. 
  \textit{Inset:} 
  The relative difference, $(F_2-F_2(R_\lambda=4141))/F_2(R_\lambda=4141)$ 
  shows universality to within $\pm3$\% (indicated by the shaded area) %of the reference curve 
  for about three decades in $r$. 
  The plots include measurements at a total of 29 different Reynolds numbers. 
  }
  \label{fig:FitResults}
  %\end{center}
\end{figure}

In Fig.~\ref{fig:FitResults} we compare the intermittency correction $\mu_{2F}$ from this model of decaying turbulence \cite{Yang2018} to an established empirical method for extracting $\mu_2$ from the data alone. 
This latter Extended Self Similarity (ESS) method was introduced in \citet{Benzi1993} and assumes that $F_n \approx F_{|3|}$, such that ratios of different order structure functions show an extended scaling range with reduced effects of the finite Reynolds number and reduced uncertainty in the inertial-range scaling exponent $\zeta_{2,ESS}$. Phenomenological models potentially connected to this empirical observation can be found in \cite{She1994,Dubrulle1994}. 
We find good agreement between this method of extended self-similarity (ESS) and the model parameter $\zeta_{2F}=\mu_{2F}+2/3$.
 
We are finally in the position to measure the universal modulation $F(R_\lambda,r/\eta)$ at large Reynolds numbers and small scales $r<0.1L$ (the statistics of large scales inevitably depend on the flow geometry). For this we consider the curve 
\begin{equation}
    F_2(R_\lambda,r/\eta) = \frac{S_2}{C_2(\varepsilon r)^{2/3}(r/\eta)^{\mu_{2}}}.
    \label{eq:DefF}
\end{equation}
We determine $C_2$ by normalising the maximum of the resulting curves to 1 and by fixing $\mu_2=0.693$ from the ESS estimate. 
Fig.~\ref{fig:FitResults} (B) shows that $F_2$ begins to collapse around $R_\lambda\approx 1500$, i.e. assumes a universal form at high $R_\lambda$. 
To show this more clearly, we take $F(R_\lambda = 4141)$ as an approximation towards this asymptotic form and plot the relative differences towards this reference. In the inset of Fig.~\ref{fig:FitResults} (A) we observe that starting around $R_\lambda\approx 1500$ the curves are within $\pm3\%$ of each other.

%%%%%%%%%
%Discussion
%%%%%%%

In this article we present experimental data that shows how the velocity increment statistics approach a fully-developed inertial range whose shape is independent of the Reynolds number. 
%Thus, the inertial range statistics are free from the direct effects of viscosity. 
While this is in agreement with Kolmogorov's hypothesis of universality, 
the scaling laws (and their intermittency corrections) anticipated for these conditions are not directly observed. 
That is, the inertial range is only approximately described by power laws and carries an apparently \emph{$R_\lambda$-independent} modulation, $F_2(r/\eta)$ in eq.~(\ref{eq:DefF}). 
% Remarkably, even drastic changes in the means of generating turbulence, such as by replacing a passive grid with an active grid, do not alter $F_2(r/\eta)$ (TODO: How do we show this?).
Data from entirely different flow geometries, such as a jet \cite{Antonia2019}, suggest that $F_n(r/\eta)$ is sensitive to the overall flow configuration for $n=3$, but less so for $n=2$. 
We observed little variation for different active grid schemes.
A careful analysis of other high $R_\lambda$-data is of great interest in the light of our results.
We also show that the widely used empirical ESS scheme to obtain the intermittency correction $\mu$  \cite{Benzi1993} gives an equivalent answer at second order to a physically motivated model of the entire structure function \cite{Yang2018}. 

In decaying turbulence the inertial range grows more slowly than in continuously forced turbulence, and the time-dependent term in the statistical evolution equation does not vanish \cite{Fukayama2000,Danaila2002,Antonia2006,Tang2017,Yang2018,Antonia2019}. 
Indeed, a model \cite{Yang2018} for the decay of turbulence (confined as in our experiment) predicts an influence of the decay on its structure from large scales to deep into the inertial range and allows us to quantify the intermittency correction $\mu_2$. 
While the model we use is designed to approach an inertial range power law, our data suggest that above $R_\lambda \approx 2000$ the approach to a power law is halted. 
We show instead that above this Reynolds number the statistics are described by a universal and nontrivial shape from small scales up to $r \approx 0.1L$. This suggests that even models that consider large-scale effects, but prescribe an asymptotic approach to K41-like scaling laws fail to describe real flows at large $R_\lambda$.

%At this point, several explanations for the observed universality without scaling appear plausible:
%The large scales, where a turbulent fluctuation was generated, might impact its  down to extremely small scales. Similarly, geometric constraints originating from viscous scales (e.g. one-dimensional vortex filaments) might have such a global impact \cite{Dubrulle1994,She1994}. 
%** i don't understand either of these plausible explanations.  GB

When turbulence is not isotropic, 
scaling laws appear only when projecting onto appropriate symmetry groups \cite{kurienAnisotropicScalingContributions2000,biferaleAnisotropyTurbulentFlows2005,Iyer2020}. 
The instrumentation in the present experiment allows only unidirectional velocity measurements, such that anisotropy can be inferred only indirectly. 
The measurements might therefore represent the approach to universality in anisotropic turbulence with little consequences for the idealised Kolmogorov framework. 
However, the measurement volume is relatively free of mean shear and the results are remarkably robust even when the turbulence is excited using an anisotropic active grid protocol or a classical and static grid. 

We have shown that $F_2$ is a $R_\lambda$-independent and nontrivial function of the scale $r$ with indications from Figs.~\ref{fig:ZetanvsR} and \ref{fig:ZetaR0vsRlam} that higher orders behave similarly. 
We point out that ESS means that $F_n$ is similar for all even orders. 
The processes that shape the asymptotic form of $F_n$ and that interfere with power-law scaling are evidently open questions. 
This already bears the potential for substantial advancements to applied turbulence models and the scaling seen in engineering wind tunnel studies. 
Future studies will need to investigate the degree to which $F_n$ changes from flow to flow at very large $R_\lambda$. While data from DNS \cite{Ishihara2020} and atmospheric measurements \cite{Tsuji2004} reproduced in Fig. \ref{LocalSlopeCollapseEta} indicate some flow-dependent variability, a demonstration of the approach towards $R_\lambda$-independence is unique to the study at hand. 
A flow-independent function of the $n$-th order statistics, if it existed, would have far-reaching implications for turbulence models and closure schemes. Moreover, a theoretical understanding of the underlying universal mechanisms would be an important step towards an efficient simulation of turbulent flows. 

To summarise, we claim that the route to universality in decaying turbulence is different from a simple removal of large-scale and viscous effects over some range of scales and the subsequent appearance of scaling laws. Past claims that this is simply an slow process \cite{Antonia2019,Tang2019,Antonia2003} to occur at extremely large $R_\lambda$ only are at odds with our data, which shows universality, but no signs of the emergence of power laws. Our data is however plausible if scale-locality is not given or if large scales directly impact significantly smaller scales (and vice versa) as suggested in Refs. \cite{Brasseur1994,Alexakis2005,Verma2007,Leung2012,Sinhuber2017}

We end by commenting that deviations from power-law scaling in the inertial range have
in the past been dismissed as finite Reynolds number effects that were to be circumvented. 
Viscous effects are important when the Reynolds number is low. 
Our results suggest however, that deviations from power-law scaling are an important feature of naturally occurring decaying turbulence, whatever its Reynolds number.

\section*{Acknowledgements}
We thank M. Hultmark and Y. Fan for providing the nanoscale hot wire probes and helping with their operation. We thank M. Sinhuber for help with the passive grid data and helpful discussions. We thank A. Pumir, H. Xu, M. Wilczek, and D. Lohse for helpful discussions. The Max Planck Variable Density Turbulence Tunnel (VDTT) is maintained and operated by A. Kubitzek, A. Kopp, and A. Renner. The machine workshop led by U. Schminke and the electronic workshop led by O. Kurre built and installed the active grid. The Max Planck Society and Volkswagen Foundation provided financial support for building the VDTT. 

\bibliography{library.bib}

%apsrev4-2.bst 2019-01-14 (MD) hand-edited version of apsrev4-1.bst
%Control: key (0)
%Control: author (8) initials jnrlst
%Control: editor formatted (1) identically to author
%Control: production of article title (0) allowed
%Control: page (0) single
%Control: year (1) truncated
%Control: production of eprint (0) enabled
\begin{thebibliography}{75}%
\makeatletter
\providecommand \@ifxundefined [1]{%
 \@ifx{#1\undefined}
}%
\providecommand \@ifnum [1]{%
 \ifnum #1\expandafter \@firstoftwo
 \else \expandafter \@secondoftwo
 \fi
}%
\providecommand \@ifx [1]{%
 \ifx #1\expandafter \@firstoftwo
 \else \expandafter \@secondoftwo
 \fi
}%
\providecommand \natexlab [1]{#1}%
\providecommand \enquote  [1]{``#1''}%
\providecommand \bibnamefont  [1]{#1}%
\providecommand \bibfnamefont [1]{#1}%
\providecommand \citenamefont [1]{#1}%
\providecommand \href@noop [0]{\@secondoftwo}%
\providecommand \href [0]{\begingroup \@sanitize@url \@href}%
\providecommand \@href[1]{\@@startlink{#1}\@@href}%
\providecommand \@@href[1]{\endgroup#1\@@endlink}%
\providecommand \@sanitize@url [0]{\catcode `\\12\catcode `\$12\catcode
  `\&12\catcode `\#12\catcode `\^12\catcode `\_12\catcode `\%12\relax}%
\providecommand \@@startlink[1]{}%
\providecommand \@@endlink[0]{}%
\providecommand \url  [0]{\begingroup\@sanitize@url \@url }%
\providecommand \@url [1]{\endgroup\@href {#1}{\urlprefix }}%
\providecommand \urlprefix  [0]{URL }%
\providecommand \Eprint [0]{\href }%
\providecommand \doibase [0]{https://doi.org/}%
\providecommand \selectlanguage [0]{\@gobble}%
\providecommand \bibinfo  [0]{\@secondoftwo}%
\providecommand \bibfield  [0]{\@secondoftwo}%
\providecommand \translation [1]{[#1]}%
\providecommand \BibitemOpen [0]{}%
\providecommand \bibitemStop [0]{}%
\providecommand \bibitemNoStop [0]{.\EOS\space}%
\providecommand \EOS [0]{\spacefactor3000\relax}%
\providecommand \BibitemShut  [1]{\csname bibitem#1\endcsname}%
\let\auto@bib@innerbib\@empty
%</preamble>
\bibitem [{\citenamefont {Bodenschatz}\ \emph
  {et~al.}(2014{\natexlab{a}})\citenamefont {Bodenschatz}, \citenamefont
  {Bewley}, \citenamefont {Nobach}, \citenamefont {Sinhuber},\ and\
  \citenamefont {Xu}}]{Bodenschatz2014}%
  \BibitemOpen
  \bibfield  {author} {\bibinfo {author} {\bibfnamefont {E.}~\bibnamefont
  {Bodenschatz}}, \bibinfo {author} {\bibfnamefont {G.~P.}\ \bibnamefont
  {Bewley}}, \bibinfo {author} {\bibfnamefont {H.}~\bibnamefont {Nobach}},
  \bibinfo {author} {\bibfnamefont {M.}~\bibnamefont {Sinhuber}},\ and\
  \bibinfo {author} {\bibfnamefont {H.}~\bibnamefont {Xu}},\ }\bibfield
  {title} {\bibinfo {title} {Variable density turbulence tunnel facility},\
  }\href {https://doi.org/10.1063/1.4896138} {\bibfield  {journal} {\bibinfo
  {journal} {Review of Scientific Instruments}\ }\textbf {\bibinfo {volume}
  {85}},\ \bibinfo {pages} {093908} (\bibinfo {year}
  {2014}{\natexlab{a}})}\BibitemShut {NoStop}%
\bibitem [{\citenamefont {Griffin}\ \emph {et~al.}(2019)\citenamefont
  {Griffin}, \citenamefont {Wei}, \citenamefont {Bodenschatz},\ and\
  \citenamefont {Bewley}}]{Griffin2019}%
  \BibitemOpen
  \bibfield  {author} {\bibinfo {author} {\bibfnamefont {K.~P.}\ \bibnamefont
  {Griffin}}, \bibinfo {author} {\bibfnamefont {N.~J.}\ \bibnamefont {Wei}},
  \bibinfo {author} {\bibfnamefont {E.}~\bibnamefont {Bodenschatz}},\ and\
  \bibinfo {author} {\bibfnamefont {G.~P.}\ \bibnamefont {Bewley}},\ }\bibfield
   {title} {\bibinfo {title} {Control of long-range correlations in
  turbulence},\ }\href {https://doi.org/10.1007/s00348-019-2698-1} {\bibfield
  {journal} {\bibinfo  {journal} {Experiments in Fluids}\ }\textbf {\bibinfo
  {volume} {60}},\ \bibinfo {pages} {55} (\bibinfo {year} {2019})},\ \Eprint
  {https://arxiv.org/abs/1809.05126} {arXiv:1809.05126} \BibitemShut {NoStop}%
\bibitem [{\citenamefont {Kolmogorov}(1941)}]{Kolmogorov1941}%
  \BibitemOpen
  \bibfield  {author} {\bibinfo {author} {\bibfnamefont {A.~N.}\ \bibnamefont
  {Kolmogorov}},\ }\bibfield  {title} {\bibinfo {title} {The {{Local
  Structure}} of {{Turbulence}} in {{Incompressible Viscous Fluid}} for {{Very
  Large Reynolds Numbers}}},\ }\href@noop {} {\bibfield  {journal} {\bibinfo
  {journal} {Proceedings: Mathematical and Physical Sciences}\ }\textbf
  {\bibinfo {volume} {434}},\ \bibinfo {pages} {9} (\bibinfo {year}
  {1941})}\BibitemShut {NoStop}%
\bibitem [{\citenamefont {Taylor}(1935)}]{Taylor1935}%
  \BibitemOpen
  \bibfield  {author} {\bibinfo {author} {\bibfnamefont {G.~I.}\ \bibnamefont
  {Taylor}},\ }\bibfield  {title} {\bibinfo {title} {Statistical theory of
  turbulenc},\ }\href {https://doi.org/10.1098/rspa.1935.0158} {\bibfield
  {journal} {\bibinfo  {journal} {Proceedings of the Royal Society A:
  Mathematical, Physical and Engineering Sciences}\ }\textbf {\bibinfo {volume}
  {151}},\ \bibinfo {pages} {421} (\bibinfo {year} {1935})}\BibitemShut
  {NoStop}%
\bibitem [{\citenamefont {Ishihara}\ \emph {et~al.}(2020)\citenamefont
  {Ishihara}, \citenamefont {Kaneda}, \citenamefont {Morishita}, \citenamefont
  {Yokokawa},\ and\ \citenamefont {Uno}}]{Ishihara2020}%
  \BibitemOpen
  \bibfield  {author} {\bibinfo {author} {\bibfnamefont {T.}~\bibnamefont
  {Ishihara}}, \bibinfo {author} {\bibfnamefont {Y.}~\bibnamefont {Kaneda}},
  \bibinfo {author} {\bibfnamefont {K.}~\bibnamefont {Morishita}}, \bibinfo
  {author} {\bibfnamefont {M.}~\bibnamefont {Yokokawa}},\ and\ \bibinfo
  {author} {\bibfnamefont {A.}~\bibnamefont {Uno}},\ }\bibfield  {title}
  {\bibinfo {title} {Second-order velocity structure functions in direct
  numerical simulations of turbulence with ${R}_{\ensuremath{\lambda}}$ up to
  2250},\ }\href {https://doi.org/10.1103/PhysRevFluids.5.104608} {\bibfield
  {journal} {\bibinfo  {journal} {Phys. Rev. Fluids}\ }\textbf {\bibinfo
  {volume} {5}},\ \bibinfo {pages} {104608} (\bibinfo {year}
  {2020})}\BibitemShut {NoStop}%
\bibitem [{\citenamefont {Tsuji}(2004{\natexlab{a}})}]{Tsuji2004}%
  \BibitemOpen
  \bibfield  {author} {\bibinfo {author} {\bibfnamefont {Y.}~\bibnamefont
  {Tsuji}},\ }\bibfield  {title} {\bibinfo {title} {Intermittency effect on
  energy spectrum in high-{{Reynolds}} number turbulence},\ }\href
  {https://doi.org/10.1063/1.1689931} {\bibfield  {journal} {\bibinfo
  {journal} {Physics of Fluids}\ }\textbf {\bibinfo {volume} {16}},\ \bibinfo
  {pages} {L43} (\bibinfo {year} {2004}{\natexlab{a}})}\BibitemShut {NoStop}%
\bibitem [{\citenamefont {Meyers}\ and\ \citenamefont
  {Baelmans}(2004)}]{Meyers2004}%
  \BibitemOpen
  \bibfield  {author} {\bibinfo {author} {\bibfnamefont {J.}~\bibnamefont
  {Meyers}}\ and\ \bibinfo {author} {\bibfnamefont {M.}~\bibnamefont
  {Baelmans}},\ }\bibfield  {title} {\bibinfo {title} {Determination of
  subfilter energy in large-eddy simulations},\ }\href
  {https://doi.org/10.1088/1468-5248/5/1/026} {\bibfield  {journal} {\bibinfo
  {journal} {Journal of Turbulence}\ }\textbf {\bibinfo {volume} {5}},\
  \bibinfo {pages} {N26} (\bibinfo {year} {2004})}\BibitemShut {NoStop}%
\bibitem [{\citenamefont {Batchelor}\ and\ \citenamefont
  {Townsend}(1947)}]{Batchelor1947}%
  \BibitemOpen
  \bibfield  {author} {\bibinfo {author} {\bibfnamefont {G.~K.}\ \bibnamefont
  {Batchelor}}\ and\ \bibinfo {author} {\bibfnamefont {A.}~\bibnamefont
  {Townsend}},\ }\bibfield  {title} {\bibinfo {title} {Decay of vorticity in
  isotropic turbulence},\ }\href {https://doi.org/10.1098/rspa.1947.0095}
  {\bibfield  {journal} {\bibinfo  {journal} {Proceedings of the Royal Society
  of London. Series A. Mathematical and Physical Sciences}\ }\textbf {\bibinfo
  {volume} {190}},\ \bibinfo {pages} {534} (\bibinfo {year}
  {1947})}\BibitemShut {NoStop}%
\bibitem [{\citenamefont {{G. K. Batchelor}}\ and\ \citenamefont {{Townsend,
  A.A.}}(1949)}]{G.K.Batchelor1949}%
  \BibitemOpen
  \bibfield  {author} {\bibinfo {author} {\bibnamefont {{G. K. Batchelor}}}\
  and\ \bibinfo {author} {\bibnamefont {{Townsend, A.A.}}},\ }\bibfield
  {title} {\bibinfo {title} {The nature of turbulent motion at large
  wave-numbers},\ }\href {https://doi.org/10.1098/rspa.1949.0136} {\bibfield
  {journal} {\bibinfo  {journal} {Proceedings of the Royal Society of London.
  Series A. Mathematical and Physical Sciences}\ }\textbf {\bibinfo {volume}
  {199}},\ \bibinfo {pages} {238} (\bibinfo {year} {1949})}\BibitemShut
  {NoStop}%
\bibitem [{\citenamefont {Kolmogorov}(1962)}]{Kolmogorov1962}%
  \BibitemOpen
  \bibfield  {author} {\bibinfo {author} {\bibfnamefont {A.~N.}\ \bibnamefont
  {Kolmogorov}},\ }\bibfield  {title} {\bibinfo {title} {A refinement of
  previous hypotheses concerning the local structure of turbulence in a viscous
  incompressible fluid at high {{Reynolds}} number},\ }\href
  {https://doi.org/10.1017/S0022112062000518} {\bibfield  {journal} {\bibinfo
  {journal} {Journal of Fluid Mechanics}\ }\textbf {\bibinfo {volume} {13}},\
  \bibinfo {pages} {82} (\bibinfo {year} {1962})}\BibitemShut {NoStop}%
\bibitem [{\citenamefont {Frisch}\ \emph {et~al.}(1978)\citenamefont {Frisch},
  \citenamefont {Sulem},\ and\ \citenamefont {Nelkin}}]{Frisch1978}%
  \BibitemOpen
  \bibfield  {author} {\bibinfo {author} {\bibfnamefont {U.}~\bibnamefont
  {Frisch}}, \bibinfo {author} {\bibfnamefont {P.-L.}\ \bibnamefont {Sulem}},\
  and\ \bibinfo {author} {\bibfnamefont {M.}~\bibnamefont {Nelkin}},\
  }\bibfield  {title} {\bibinfo {title} {A simple dynamical model of
  intermittent fully developed turbulence},\ }\href
  {https://doi.org/10.1017/S0022112078001846} {\bibfield  {journal} {\bibinfo
  {journal} {Journal of Fluid Mechanics}\ }\textbf {\bibinfo {volume} {87}},\
  \bibinfo {pages} {719} (\bibinfo {year} {1978})}\BibitemShut {NoStop}%
\bibitem [{\citenamefont {Benzi}\ \emph {et~al.}(1984)\citenamefont {Benzi},
  \citenamefont {Paladin}, \citenamefont {Parisi},\ and\ \citenamefont
  {Vulpiani}}]{Benzi1984}%
  \BibitemOpen
  \bibfield  {author} {\bibinfo {author} {\bibfnamefont {R.}~\bibnamefont
  {Benzi}}, \bibinfo {author} {\bibfnamefont {G.}~\bibnamefont {Paladin}},
  \bibinfo {author} {\bibfnamefont {G.}~\bibnamefont {Parisi}},\ and\ \bibinfo
  {author} {\bibfnamefont {A.}~\bibnamefont {Vulpiani}},\ }\bibfield  {title}
  {\bibinfo {title} {On the multifractal nature of fully developed turbulence
  and chaotic systems},\ }\href {https://doi.org/10.1088/0305-4470/17/18/021}
  {\bibfield  {journal} {\bibinfo  {journal} {Journal of Physics A:
  Mathematical and General}\ }\textbf {\bibinfo {volume} {17}},\ \bibinfo
  {pages} {3521} (\bibinfo {year} {1984})}\BibitemShut {NoStop}%
\bibitem [{\citenamefont {Sreenivasan}\ and\ \citenamefont
  {Meneveau}(1986)}]{Sreenivasan1986}%
  \BibitemOpen
  \bibfield  {author} {\bibinfo {author} {\bibfnamefont {K.~R.}\ \bibnamefont
  {Sreenivasan}}\ and\ \bibinfo {author} {\bibfnamefont {C.}~\bibnamefont
  {Meneveau}},\ }\bibfield  {title} {\bibinfo {title} {The fractal facets of
  turbulence},\ }\href {https://doi.org/10.1017/S0022112086001209} {\bibfield
  {journal} {\bibinfo  {journal} {Journal of Fluid Mechanics}\ }\textbf
  {\bibinfo {volume} {173}},\ \bibinfo {pages} {357} (\bibinfo {year}
  {1986})}\BibitemShut {NoStop}%
\bibitem [{\citenamefont {Meneveau}\ and\ \citenamefont
  {Sreenivasan}(1987)}]{Meneveau1987}%
  \BibitemOpen
  \bibfield  {author} {\bibinfo {author} {\bibfnamefont {C.}~\bibnamefont
  {Meneveau}}\ and\ \bibinfo {author} {\bibfnamefont {K.~R.}\ \bibnamefont
  {Sreenivasan}},\ }\bibfield  {title} {\bibinfo {title} {Simple multifractal
  cascade model for fully developed turbulence},\ }\href
  {https://doi.org/10.1103/PhysRevLett.59.1424} {\bibfield  {journal} {\bibinfo
   {journal} {Physical Review Letters}\ }\textbf {\bibinfo {volume} {59}},\
  \bibinfo {pages} {1424} (\bibinfo {year} {1987})}\BibitemShut {NoStop}%
\bibitem [{\citenamefont {Andrews}\ \emph {et~al.}(1989)\citenamefont
  {Andrews}, \citenamefont {Phillips}, \citenamefont {Shivamoggi},
  \citenamefont {Beck},\ and\ \citenamefont
  {Joshi}}]{andrewsStatisticalTheoryDistribution1989}%
  \BibitemOpen
  \bibfield  {author} {\bibinfo {author} {\bibfnamefont {L.~C.}\ \bibnamefont
  {Andrews}}, \bibinfo {author} {\bibfnamefont {R.~L.}\ \bibnamefont
  {Phillips}}, \bibinfo {author} {\bibfnamefont {B.~K.}\ \bibnamefont
  {Shivamoggi}}, \bibinfo {author} {\bibfnamefont {J.~K.}\ \bibnamefont
  {Beck}},\ and\ \bibinfo {author} {\bibfnamefont {M.~L.}\ \bibnamefont
  {Joshi}},\ }\bibfield  {title} {\bibinfo {title} {A statistical theory for
  the distribution of energy dissipation in intermittent turbulence},\ }\href
  {https://doi.org/10/c27qpd} {\bibfield  {journal} {\bibinfo  {journal}
  {Physics of Fluids A: Fluid Dynamics}\ }\textbf {\bibinfo {volume} {1}},\
  \bibinfo {pages} {999} (\bibinfo {year} {1989})}\BibitemShut {NoStop}%
\bibitem [{\citenamefont {She}\ and\ \citenamefont {Leveque}(1994)}]{She1994}%
  \BibitemOpen
  \bibfield  {author} {\bibinfo {author} {\bibfnamefont {Z.-S.}\ \bibnamefont
  {She}}\ and\ \bibinfo {author} {\bibfnamefont {E.}~\bibnamefont {Leveque}},\
  }\bibfield  {title} {\bibinfo {title} {Universal scaling laws in fully
  developed turbulence},\ }\href {https://doi.org/10.1103/PhysRevLett.72.336}
  {\bibfield  {journal} {\bibinfo  {journal} {Physical Review Letters}\
  }\textbf {\bibinfo {volume} {72}},\ \bibinfo {pages} {336} (\bibinfo {year}
  {1994})}\BibitemShut {NoStop}%
\bibitem [{\citenamefont {Dubrulle}(1994)}]{Dubrulle1994}%
  \BibitemOpen
  \bibfield  {author} {\bibinfo {author} {\bibfnamefont {B.}~\bibnamefont
  {Dubrulle}},\ }\bibfield  {title} {\bibinfo {title} {Intermittency in fully
  developed turbulence: {{Log}}-{{Poisson}} statistics and generalized scale
  covariance},\ }\href {https://doi.org/10.1103/PhysRevLett.73.959} {\bibfield
  {journal} {\bibinfo  {journal} {Physical Review Letters}\ }\textbf {\bibinfo
  {volume} {73}},\ \bibinfo {pages} {959} (\bibinfo {year} {1994})}\BibitemShut
  {NoStop}%
\bibitem [{\citenamefont {Barenblatt}\ and\ \citenamefont
  {Goldenfeld}(1995)}]{Barenblatt1995}%
  \BibitemOpen
  \bibfield  {author} {\bibinfo {author} {\bibfnamefont {G.~I.}\ \bibnamefont
  {Barenblatt}}\ and\ \bibinfo {author} {\bibfnamefont {N.}~\bibnamefont
  {Goldenfeld}},\ }\bibfield  {title} {\bibinfo {title} {Does fully developed
  turbulence exist? {{Reynolds}} number independence versus asymptotic
  covariance},\ }\href {https://doi.org/10.1063/1.868685} {\bibfield  {journal}
  {\bibinfo  {journal} {Physics of Fluids}\ }\textbf {\bibinfo {volume} {7}},\
  \bibinfo {pages} {3078} (\bibinfo {year} {1995})}\BibitemShut {NoStop}%
\bibitem [{\citenamefont {Praskovsky}\ and\ \citenamefont
  {Oncley}(1994)}]{Praskovsky1994}%
  \BibitemOpen
  \bibfield  {author} {\bibinfo {author} {\bibfnamefont {A.}~\bibnamefont
  {Praskovsky}}\ and\ \bibinfo {author} {\bibfnamefont {S.}~\bibnamefont
  {Oncley}},\ }\bibfield  {title} {\bibinfo {title} {Measurements of the
  {{Kolmogorov}} constant and intermittency exponent at very high {{Reynolds}}
  numbers},\ }\href {https://doi.org/10.1063/1.868435} {\bibfield  {journal}
  {\bibinfo  {journal} {Physics of Fluids}\ }\textbf {\bibinfo {volume} {6}},\
  \bibinfo {pages} {2886} (\bibinfo {year} {1994})}\BibitemShut {NoStop}%
\bibitem [{\citenamefont {Sreenivasan}(1998)}]{Sreenivasan1998}%
  \BibitemOpen
  \bibfield  {author} {\bibinfo {author} {\bibfnamefont {K.~R.}\ \bibnamefont
  {Sreenivasan}},\ }\bibfield  {title} {\bibinfo {title} {An update on the
  energy dissipation rate in isotropic turbulence},\ }\href
  {https://doi.org/10.1063/1.869575} {\bibfield  {journal} {\bibinfo  {journal}
  {Physics of Fluids}\ }\textbf {\bibinfo {volume} {10}},\ \bibinfo {pages}
  {528} (\bibinfo {year} {1998})}\BibitemShut {NoStop}%
\bibitem [{\citenamefont {Kahalerras}\ \emph {et~al.}(1998)\citenamefont
  {Kahalerras}, \citenamefont {Mal{\'e}cot}, \citenamefont {Gagne},\ and\
  \citenamefont {Castaing}}]{Kahalerras1998}%
  \BibitemOpen
  \bibfield  {author} {\bibinfo {author} {\bibfnamefont {H.}~\bibnamefont
  {Kahalerras}}, \bibinfo {author} {\bibfnamefont {Y.}~\bibnamefont
  {Mal{\'e}cot}}, \bibinfo {author} {\bibfnamefont {Y.}~\bibnamefont {Gagne}},\
  and\ \bibinfo {author} {\bibfnamefont {B.}~\bibnamefont {Castaing}},\
  }\bibfield  {title} {\bibinfo {title} {Intermittency and {{Reynolds}}
  number},\ }\href {https://doi.org/10.1063/1.869613} {\bibfield  {journal}
  {\bibinfo  {journal} {Physics of Fluids}\ }\textbf {\bibinfo {volume} {10}},\
  \bibinfo {pages} {910} (\bibinfo {year} {1998})}\BibitemShut {NoStop}%
\bibitem [{\citenamefont {Tsuji}(2004{\natexlab{b}})}]{Tsuji2004a}%
  \BibitemOpen
  \bibfield  {author} {\bibinfo {author} {\bibfnamefont {Y.}~\bibnamefont
  {Tsuji}},\ }\bibfield  {title} {\bibinfo {title} {Intermittency effect on
  energy spectrum in high-{{Reynolds}} number turbulence},\ }\href
  {https://doi.org/10.1063/1.1689931} {\bibfield  {journal} {\bibinfo
  {journal} {Physics of Fluids}\ }\textbf {\bibinfo {volume} {16}},\ \bibinfo
  {pages} {L43} (\bibinfo {year} {2004}{\natexlab{b}})}\BibitemShut {NoStop}%
\bibitem [{\citenamefont {White}\ \emph {et~al.}(2002)\citenamefont {White},
  \citenamefont {Karpetis},\ and\ \citenamefont {Sreenivasan}}]{White2002}%
  \BibitemOpen
  \bibfield  {author} {\bibinfo {author} {\bibfnamefont {C.~M.}\ \bibnamefont
  {White}}, \bibinfo {author} {\bibfnamefont {A.~N.}\ \bibnamefont
  {Karpetis}},\ and\ \bibinfo {author} {\bibfnamefont {K.~R.}\ \bibnamefont
  {Sreenivasan}},\ }\bibfield  {title} {\bibinfo {title}
  {High-{{Reynolds}}-number turbulence in small apparatus: Grid turbulence in
  cryogenic liquids},\ }\href {https://doi.org/10.1017/S0022112001007194}
  {\bibfield  {journal} {\bibinfo  {journal} {Journal of Fluid Mechanics}\
  }\textbf {\bibinfo {volume} {452}},\ \bibinfo {pages} {189} (\bibinfo {year}
  {2002})}\BibitemShut {NoStop}%
\bibitem [{\citenamefont {Pietropinto}\ \emph {et~al.}(2003)\citenamefont
  {Pietropinto}, \citenamefont {Poulain}, \citenamefont {Baudet}, \citenamefont
  {Castaing}, \citenamefont {Chabaud}, \citenamefont {Gagne}, \citenamefont
  {H{\'e}bral}, \citenamefont {Ladam}, \citenamefont {Lebrun}, \citenamefont
  {Pirotte},\ and\ \citenamefont {Roche}}]{Pietropinto2003}%
  \BibitemOpen
  \bibfield  {author} {\bibinfo {author} {\bibfnamefont {S.}~\bibnamefont
  {Pietropinto}}, \bibinfo {author} {\bibfnamefont {C.}~\bibnamefont
  {Poulain}}, \bibinfo {author} {\bibfnamefont {C.}~\bibnamefont {Baudet}},
  \bibinfo {author} {\bibfnamefont {B.}~\bibnamefont {Castaing}}, \bibinfo
  {author} {\bibfnamefont {B.}~\bibnamefont {Chabaud}}, \bibinfo {author}
  {\bibfnamefont {Y.}~\bibnamefont {Gagne}}, \bibinfo {author} {\bibfnamefont
  {B.}~\bibnamefont {H{\'e}bral}}, \bibinfo {author} {\bibfnamefont
  {Y.}~\bibnamefont {Ladam}}, \bibinfo {author} {\bibfnamefont
  {P.}~\bibnamefont {Lebrun}}, \bibinfo {author} {\bibfnamefont
  {O.}~\bibnamefont {Pirotte}},\ and\ \bibinfo {author} {\bibfnamefont
  {P.}~\bibnamefont {Roche}},\ }\bibfield  {title} {\bibinfo {title}
  {Superconducting instrumentation for high {{Reynolds}} turbulence experiments
  with low temperature gaseous helium},\ }\href
  {https://doi.org/10.1016/S0921-4534(02)02115-9} {\bibfield  {journal}
  {\bibinfo  {journal} {Physica C: Superconductivity}\ }\textbf {\bibinfo
  {volume} {386}},\ \bibinfo {pages} {512} (\bibinfo {year}
  {2003})}\BibitemShut {NoStop}%
\bibitem [{\citenamefont {Bewley}\ and\ \citenamefont
  {Sreenivasan}(2009)}]{Bewley2009}%
  \BibitemOpen
  \bibfield  {author} {\bibinfo {author} {\bibfnamefont {G.~P.}\ \bibnamefont
  {Bewley}}\ and\ \bibinfo {author} {\bibfnamefont {K.~R.}\ \bibnamefont
  {Sreenivasan}},\ }\bibfield  {title} {\bibinfo {title} {The {{Decay}} of a
  {{Quantized Vortex Ring}} and the {{Influence}} of {{Tracer Particles}}},\
  }\href {https://doi.org/10.1007/s10909-009-9903-1} {\bibfield  {journal}
  {\bibinfo  {journal} {Journal of Low Temperature Physics}\ }\textbf {\bibinfo
  {volume} {156}},\ \bibinfo {pages} {84} (\bibinfo {year} {2009})}\BibitemShut
  {NoStop}%
\bibitem [{\citenamefont {Salort}\ \emph {et~al.}(2012)\citenamefont {Salort},
  \citenamefont {Chabaud}, \citenamefont {L{\'e}v{\^e}que},\ and\ \citenamefont
  {Roche}}]{Salort2012}%
  \BibitemOpen
  \bibfield  {author} {\bibinfo {author} {\bibfnamefont {J.}~\bibnamefont
  {Salort}}, \bibinfo {author} {\bibfnamefont {B.}~\bibnamefont {Chabaud}},
  \bibinfo {author} {\bibfnamefont {E.}~\bibnamefont {L{\'e}v{\^e}que}},\ and\
  \bibinfo {author} {\bibfnamefont {P.-E.}\ \bibnamefont {Roche}},\ }\bibfield
  {title} {\bibinfo {title} {Energy cascade and the four-fifths law in
  superfluid turbulence},\ }\href {https://doi.org/10.1209/0295-5075/97/34006}
  {\bibfield  {journal} {\bibinfo  {journal} {EPL (Europhysics Letters)}\
  }\textbf {\bibinfo {volume} {97}},\ \bibinfo {pages} {34006} (\bibinfo {year}
  {2012})}\BibitemShut {NoStop}%
\bibitem [{\citenamefont {Rousset}\ \emph {et~al.}(2014)\citenamefont
  {Rousset}, \citenamefont {Bonnay}, \citenamefont {Diribarne}, \citenamefont
  {Girard}, \citenamefont {Poncet}, \citenamefont {Herbert}, \citenamefont
  {Salort}, \citenamefont {Baudet}, \citenamefont {Castaing}, \citenamefont
  {Chevillard}, \citenamefont {Daviaud}, \citenamefont {Dubrulle},
  \citenamefont {Gagne}, \citenamefont {Gibert}, \citenamefont {H{\'e}bral},
  \citenamefont {Lehner}, \citenamefont {Roche}, \citenamefont
  {{Saint-Michel}},\ and\ \citenamefont {Bon~Mardion}}]{Rousset2014}%
  \BibitemOpen
  \bibfield  {author} {\bibinfo {author} {\bibfnamefont {B.}~\bibnamefont
  {Rousset}}, \bibinfo {author} {\bibfnamefont {P.}~\bibnamefont {Bonnay}},
  \bibinfo {author} {\bibfnamefont {P.}~\bibnamefont {Diribarne}}, \bibinfo
  {author} {\bibfnamefont {A.}~\bibnamefont {Girard}}, \bibinfo {author}
  {\bibfnamefont {J.~M.}\ \bibnamefont {Poncet}}, \bibinfo {author}
  {\bibfnamefont {E.}~\bibnamefont {Herbert}}, \bibinfo {author} {\bibfnamefont
  {J.}~\bibnamefont {Salort}}, \bibinfo {author} {\bibfnamefont
  {C.}~\bibnamefont {Baudet}}, \bibinfo {author} {\bibfnamefont
  {B.}~\bibnamefont {Castaing}}, \bibinfo {author} {\bibfnamefont
  {L.}~\bibnamefont {Chevillard}}, \bibinfo {author} {\bibfnamefont
  {F.}~\bibnamefont {Daviaud}}, \bibinfo {author} {\bibfnamefont
  {B.}~\bibnamefont {Dubrulle}}, \bibinfo {author} {\bibfnamefont
  {Y.}~\bibnamefont {Gagne}}, \bibinfo {author} {\bibfnamefont
  {M.}~\bibnamefont {Gibert}}, \bibinfo {author} {\bibfnamefont
  {B.}~\bibnamefont {H{\'e}bral}}, \bibinfo {author} {\bibfnamefont
  {T.}~\bibnamefont {Lehner}}, \bibinfo {author} {\bibfnamefont {P.-E.}\
  \bibnamefont {Roche}}, \bibinfo {author} {\bibfnamefont {B.}~\bibnamefont
  {{Saint-Michel}}},\ and\ \bibinfo {author} {\bibfnamefont {M.}~\bibnamefont
  {Bon~Mardion}},\ }\bibfield  {title} {\bibinfo {title} {Superfluid high
  {{REynolds}} von {{K\'arm\'an}} experiment},\ }\href
  {https://doi.org/10.1063/1.4897542} {\bibfield  {journal} {\bibinfo
  {journal} {Review of Scientific Instruments}\ }\textbf {\bibinfo {volume}
  {85}},\ \bibinfo {pages} {103908} (\bibinfo {year} {2014})}\BibitemShut
  {NoStop}%
\bibitem [{\citenamefont {Yang}\ \emph {et~al.}(2018)\citenamefont {Yang},
  \citenamefont {Pumir},\ and\ \citenamefont {Xu}}]{Yang2018}%
  \BibitemOpen
  \bibfield  {author} {\bibinfo {author} {\bibfnamefont {P.-F.}\ \bibnamefont
  {Yang}}, \bibinfo {author} {\bibfnamefont {A.}~\bibnamefont {Pumir}},\ and\
  \bibinfo {author} {\bibfnamefont {H.}~\bibnamefont {Xu}},\ }\bibfield
  {title} {\bibinfo {title} {Generalized self-similar spectrum and the effect
  of large-scale in decaying homogeneous isotropic turbulence},\ }\href
  {https://doi.org/10.1088/1367-2630/aae72d} {\bibfield  {journal} {\bibinfo
  {journal} {New Journal of Physics}\ }\textbf {\bibinfo {volume} {20}},\
  \bibinfo {pages} {103035} (\bibinfo {year} {2018})}\BibitemShut {NoStop}%
\bibitem [{\citenamefont {Saddoughi}\ and\ \citenamefont
  {Veeravalli}(1994)}]{Saddoughi1994}%
  \BibitemOpen
  \bibfield  {author} {\bibinfo {author} {\bibfnamefont {S.~G.}\ \bibnamefont
  {Saddoughi}}\ and\ \bibinfo {author} {\bibfnamefont {S.~V.}\ \bibnamefont
  {Veeravalli}},\ }\bibfield  {title} {\bibinfo {title} {Local isotropy in
  turbulent boundary layers at high {{Reynolds}} number},\ }\href
  {https://doi.org/10.1017/S0022112094001370} {\bibfield  {journal} {\bibinfo
  {journal} {Journal of Fluid Mechanics}\ }\textbf {\bibinfo {volume} {268}},\
  \bibinfo {pages} {333} (\bibinfo {year} {1994})}\BibitemShut {NoStop}%
\bibitem [{\citenamefont {Mydlarski}\ and\ \citenamefont
  {Warhaft}(1996)}]{Mydlarski1996}%
  \BibitemOpen
  \bibfield  {author} {\bibinfo {author} {\bibfnamefont {L.}~\bibnamefont
  {Mydlarski}}\ and\ \bibinfo {author} {\bibfnamefont {Z.}~\bibnamefont
  {Warhaft}},\ }\bibfield  {title} {\bibinfo {title} {On the onset of
  high-{{Reynolds}}-number grid-generated wind tunnel turbulence},\ }\href
  {https://doi.org/10.1017/S0022112096007562} {\bibfield  {journal} {\bibinfo
  {journal} {Journal of Fluid Mechanics}\ }\textbf {\bibinfo {volume} {320}},\
  \bibinfo {pages} {331} (\bibinfo {year} {1996})}\BibitemShut {NoStop}%
\bibitem [{\citenamefont {Antonia}\ \emph {et~al.}(2019)\citenamefont
  {Antonia}, \citenamefont {Tang}, \citenamefont {Djenidi},\ and\ \citenamefont
  {Zhou}}]{Antonia2019}%
  \BibitemOpen
  \bibfield  {author} {\bibinfo {author} {\bibfnamefont {R.~A.}\ \bibnamefont
  {Antonia}}, \bibinfo {author} {\bibfnamefont {S.~L.}\ \bibnamefont {Tang}},
  \bibinfo {author} {\bibfnamefont {L.}~\bibnamefont {Djenidi}},\ and\ \bibinfo
  {author} {\bibfnamefont {Y.}~\bibnamefont {Zhou}},\ }\bibfield  {title}
  {\bibinfo {title} {Finite {{Reynolds}} number effect and the 4/5 law},\
  }\href {https://doi.org/10.1103/PhysRevFluids.4.084602} {\bibfield  {journal}
  {\bibinfo  {journal} {Physical Review Fluids}\ }\textbf {\bibinfo {volume}
  {4}},\ \bibinfo {pages} {084602} (\bibinfo {year} {2019})}\BibitemShut
  {NoStop}%
\bibitem [{\citenamefont {Sinhuber}\ \emph {et~al.}(2017)\citenamefont
  {Sinhuber}, \citenamefont {Bewley},\ and\ \citenamefont
  {Bodenschatz}}]{Sinhuber2017}%
  \BibitemOpen
  \bibfield  {author} {\bibinfo {author} {\bibfnamefont {M.}~\bibnamefont
  {Sinhuber}}, \bibinfo {author} {\bibfnamefont {G.~P.}\ \bibnamefont
  {Bewley}},\ and\ \bibinfo {author} {\bibfnamefont {E.}~\bibnamefont
  {Bodenschatz}},\ }\bibfield  {title} {\bibinfo {title} {Dissipative
  {{Effects}} on {{Inertial}}-{{Range Statistics}} at {{High Reynolds
  Numbers}}},\ }\href {https://doi.org/10.1103/PhysRevLett.119.134502}
  {\bibfield  {journal} {\bibinfo  {journal} {Physical Review Letters}\
  }\textbf {\bibinfo {volume} {119}},\ \bibinfo {pages} {134502} (\bibinfo
  {year} {2017})}\BibitemShut {NoStop}%
\bibitem [{\citenamefont {K{\"u}chler}\ \emph {et~al.}(2019)\citenamefont
  {K{\"u}chler}, \citenamefont {Bewley},\ and\ \citenamefont
  {Bodenschatz}}]{Kuchler2019}%
  \BibitemOpen
  \bibfield  {author} {\bibinfo {author} {\bibfnamefont {C.}~\bibnamefont
  {K{\"u}chler}}, \bibinfo {author} {\bibfnamefont {G.~P.}\ \bibnamefont
  {Bewley}},\ and\ \bibinfo {author} {\bibfnamefont {E.}~\bibnamefont
  {Bodenschatz}},\ }\bibfield  {title} {\bibinfo {title} {Experimental
  {{Study}} of the {{Bottleneck}} in {{Fully Developed Turbulence}}},\ }\href
  {https://doi.org/10.1007/s10955-019-02251-1} {\bibfield  {journal} {\bibinfo
  {journal} {Journal of Statistical Physics}\ }\textbf {\bibinfo {volume}
  {175}},\ \bibinfo {pages} {617} (\bibinfo {year} {2019})},\ \Eprint
  {https://arxiv.org/abs/1812.01370} {arXiv:1812.01370} \BibitemShut {NoStop}%
\bibitem [{\citenamefont {Fukayama}\ \emph {et~al.}(2000)\citenamefont
  {Fukayama}, \citenamefont {Oyamada}, \citenamefont {Nakano}, \citenamefont
  {Gotoh},\ and\ \citenamefont {Yamamoto}}]{Fukayama2000}%
  \BibitemOpen
  \bibfield  {author} {\bibinfo {author} {\bibfnamefont {D.}~\bibnamefont
  {Fukayama}}, \bibinfo {author} {\bibfnamefont {T.}~\bibnamefont {Oyamada}},
  \bibinfo {author} {\bibfnamefont {T.}~\bibnamefont {Nakano}}, \bibinfo
  {author} {\bibfnamefont {T.}~\bibnamefont {Gotoh}},\ and\ \bibinfo {author}
  {\bibfnamefont {K.}~\bibnamefont {Yamamoto}},\ }\bibfield  {title} {\bibinfo
  {title} {Longitudinal {{Structure Functions}} in {{Decaying}} and {{Forced
  Turbulence}}},\ }\href {https://doi.org/10.1143/JPSJ.69.701} {\bibfield
  {journal} {\bibinfo  {journal} {Journal of the Physical Society of Japan}\
  }\textbf {\bibinfo {volume} {69}},\ \bibinfo {pages} {701} (\bibinfo {year}
  {2000})},\ \Eprint {https://arxiv.org/abs/chao-dyn/9912033}
  {arXiv:chao-dyn/9912033} \BibitemShut {NoStop}%
\bibitem [{\citenamefont {Gotoh}\ \emph {et~al.}(2002)\citenamefont {Gotoh},
  \citenamefont {Fukayama},\ and\ \citenamefont {Nakano}}]{Gotoh2002}%
  \BibitemOpen
  \bibfield  {author} {\bibinfo {author} {\bibfnamefont {T.}~\bibnamefont
  {Gotoh}}, \bibinfo {author} {\bibfnamefont {D.}~\bibnamefont {Fukayama}},\
  and\ \bibinfo {author} {\bibfnamefont {T.}~\bibnamefont {Nakano}},\
  }\bibfield  {title} {\bibinfo {title} {Velocity field statistics in
  homogeneous steady turbulence obtained using a high-resolution direct
  numerical simulation},\ }\href {https://doi.org/10.1063/1.1448296} {\bibfield
   {journal} {\bibinfo  {journal} {Physics of Fluids}\ }\textbf {\bibinfo
  {volume} {14}},\ \bibinfo {pages} {1065} (\bibinfo {year}
  {2002})}\BibitemShut {NoStop}%
\bibitem [{\citenamefont {Chen}\ \emph {et~al.}(2005)\citenamefont {Chen},
  \citenamefont {Dhruva}, \citenamefont {Kurien}, \citenamefont {Sreenivasan},\
  and\ \citenamefont {Taylor}}]{CHEN2005}%
  \BibitemOpen
  \bibfield  {author} {\bibinfo {author} {\bibfnamefont {S.~Y.}\ \bibnamefont
  {Chen}}, \bibinfo {author} {\bibfnamefont {B.}~\bibnamefont {Dhruva}},
  \bibinfo {author} {\bibfnamefont {S.}~\bibnamefont {Kurien}}, \bibinfo
  {author} {\bibfnamefont {K.~R.}\ \bibnamefont {Sreenivasan}},\ and\ \bibinfo
  {author} {\bibfnamefont {M.~A.}\ \bibnamefont {Taylor}},\ }\bibfield  {title}
  {\bibinfo {title} {Anomalous scaling of low-order structure functions of
  turbulent velocity},\ }\bibfield  {journal} {\bibinfo  {journal} {Journal of
  Fluid Mechanics}\ }\textbf {\bibinfo {volume} {533}},\ \href
  {https://doi.org/10.1017/S002211200500443X} {10.1017/S002211200500443X}
  (\bibinfo {year} {2005})\BibitemShut {NoStop}%
\bibitem [{\citenamefont {Tang}\ \emph {et~al.}(2017)\citenamefont {Tang},
  \citenamefont {Antonia}, \citenamefont {Djenidi}, \citenamefont {Danaila},\
  and\ \citenamefont {Zhou}}]{Tang2017}%
  \BibitemOpen
  \bibfield  {author} {\bibinfo {author} {\bibfnamefont {S.~L.}\ \bibnamefont
  {Tang}}, \bibinfo {author} {\bibfnamefont {R.~A.}\ \bibnamefont {Antonia}},
  \bibinfo {author} {\bibfnamefont {L.}~\bibnamefont {Djenidi}}, \bibinfo
  {author} {\bibfnamefont {L.}~\bibnamefont {Danaila}},\ and\ \bibinfo {author}
  {\bibfnamefont {Y.}~\bibnamefont {Zhou}},\ }\bibfield  {title} {\bibinfo
  {title} {Finite {{Reynolds}} number effect on the scaling range behaviour of
  turbulent longitudinal velocity structure functions},\ }\href
  {https://doi.org/10.1017/jfm.2017.218} {\bibfield  {journal} {\bibinfo
  {journal} {Journal of Fluid Mechanics}\ }\textbf {\bibinfo {volume} {820}},\
  \bibinfo {pages} {341} (\bibinfo {year} {2017})}\BibitemShut {NoStop}%
\bibitem [{\citenamefont {Yeung}\ \emph {et~al.}(2018)\citenamefont {Yeung},
  \citenamefont {Sreenivasan},\ and\ \citenamefont {Pope}}]{Yeung2018}%
  \BibitemOpen
  \bibfield  {author} {\bibinfo {author} {\bibfnamefont {P.~K.}\ \bibnamefont
  {Yeung}}, \bibinfo {author} {\bibfnamefont {K.~R.}\ \bibnamefont
  {Sreenivasan}},\ and\ \bibinfo {author} {\bibfnamefont {S.~B.}\ \bibnamefont
  {Pope}},\ }\bibfield  {title} {\bibinfo {title} {Effects of finite spatial
  and temporal resolution in direct numerical simulations of incompressible
  isotropic turbulence},\ }\href
  {https://doi.org/10.1103/PhysRevFluids.3.064603} {\bibfield  {journal}
  {\bibinfo  {journal} {Physical Review Fluids}\ }\textbf {\bibinfo {volume}
  {3}},\ \bibinfo {pages} {064603} (\bibinfo {year} {2018})}\BibitemShut
  {NoStop}%
\bibitem [{\citenamefont {Danaila}\ \emph {et~al.}(2002)\citenamefont
  {Danaila}, \citenamefont {Anselmet},\ and\ \citenamefont
  {Antonia}}]{Danaila2002}%
  \BibitemOpen
  \bibfield  {author} {\bibinfo {author} {\bibfnamefont {L.}~\bibnamefont
  {Danaila}}, \bibinfo {author} {\bibfnamefont {F.}~\bibnamefont {Anselmet}},\
  and\ \bibinfo {author} {\bibfnamefont {R.~A.}\ \bibnamefont {Antonia}},\
  }\bibfield  {title} {\bibinfo {title} {An overview of the effect of
  large-scale inhomogeneities on small-scale turbulence},\ }\href
  {https://doi.org/10.1063/1.1476300} {\bibfield  {journal} {\bibinfo
  {journal} {Physics of Fluids}\ }\textbf {\bibinfo {volume} {14}},\ \bibinfo
  {pages} {2475} (\bibinfo {year} {2002})}\BibitemShut {NoStop}%
\bibitem [{\citenamefont {Antonia}\ \emph {et~al.}(2015)\citenamefont
  {Antonia}, \citenamefont {Tang}, \citenamefont {Djenidi},\ and\ \citenamefont
  {Danaila}}]{Antonia2015}%
  \BibitemOpen
  \bibfield  {author} {\bibinfo {author} {\bibfnamefont {R.~A.}\ \bibnamefont
  {Antonia}}, \bibinfo {author} {\bibfnamefont {S.~L.}\ \bibnamefont {Tang}},
  \bibinfo {author} {\bibfnamefont {L.}~\bibnamefont {Djenidi}},\ and\ \bibinfo
  {author} {\bibfnamefont {L.}~\bibnamefont {Danaila}},\ }\bibfield  {title}
  {\bibinfo {title} {Boundedness of the velocity derivative skewness in various
  turbulent flows},\ }\href {https://doi.org/10.1017/jfm.2015.539} {\bibfield
  {journal} {\bibinfo  {journal} {Journal of Fluid Mechanics}\ }\textbf
  {\bibinfo {volume} {781}},\ \bibinfo {pages} {727} (\bibinfo {year}
  {2015})}\BibitemShut {NoStop}%
\bibitem [{\citenamefont {Bos}\ \emph {et~al.}(2012)\citenamefont {Bos},
  \citenamefont {Chevillard}, \citenamefont {Scott},\ and\ \citenamefont
  {Rubinstein}}]{Bos2012}%
  \BibitemOpen
  \bibfield  {author} {\bibinfo {author} {\bibfnamefont {W.~J.~T.}\
  \bibnamefont {Bos}}, \bibinfo {author} {\bibfnamefont {L.}~\bibnamefont
  {Chevillard}}, \bibinfo {author} {\bibfnamefont {J.~F.}\ \bibnamefont
  {Scott}},\ and\ \bibinfo {author} {\bibfnamefont {R.}~\bibnamefont
  {Rubinstein}},\ }\bibfield  {title} {\bibinfo {title} {Reynolds number effect
  on the velocity increment skewness in isotropic turbulence},\ }\href
  {https://doi.org/10.1063/1.3678338} {\bibfield  {journal} {\bibinfo
  {journal} {Physics of Fluids}\ }\textbf {\bibinfo {volume} {24}},\ \bibinfo
  {pages} {015108} (\bibinfo {year} {2012})}\BibitemShut {NoStop}%
\bibitem [{\citenamefont {de~Karman}\ and\ \citenamefont
  {Howarth}()}]{deKarman1938}%
  \BibitemOpen
  \bibfield  {author} {\bibinfo {author} {\bibfnamefont {T.}~\bibnamefont
  {de~Karman}}\ and\ \bibinfo {author} {\bibfnamefont {L.}~\bibnamefont
  {Howarth}},\ }\bibfield  {title} {\bibinfo {title} {On the {{Statistical
  Theory}} of {{Isotropic Turbulence}}},\ }\href
  {https://doi.org/10.1098/rspa.1938.0013} {\ \textbf {\bibinfo {volume}
  {164}},\ \bibinfo {pages} {192}}\BibitemShut {NoStop}%
\bibitem [{\citenamefont {Antonia}\ \emph {et~al.}(2003)\citenamefont
  {Antonia}, \citenamefont {Smalley}, \citenamefont {Zhou}, \citenamefont
  {Anselmet},\ and\ \citenamefont {Danaila}}]{Antonia2003}%
  \BibitemOpen
  \bibfield  {author} {\bibinfo {author} {\bibfnamefont {R.~A.}\ \bibnamefont
  {Antonia}}, \bibinfo {author} {\bibfnamefont {R.~J.}\ \bibnamefont
  {Smalley}}, \bibinfo {author} {\bibfnamefont {T.}~\bibnamefont {Zhou}},
  \bibinfo {author} {\bibfnamefont {F.}~\bibnamefont {Anselmet}},\ and\
  \bibinfo {author} {\bibfnamefont {L.}~\bibnamefont {Danaila}},\ }\bibfield
  {title} {\bibinfo {title} {Similarity of energy structure functions in
  decaying homogeneous isotropic turbulence},\ }\href
  {https://doi.org/10.1017/S0022112003004713} {\bibfield  {journal} {\bibinfo
  {journal} {Journal of Fluid Mechanics}\ }\textbf {\bibinfo {volume} {487}},\
  \bibinfo {pages} {245} (\bibinfo {year} {2003})}\BibitemShut {NoStop}%
\bibitem [{\citenamefont {Danaila}\ \emph {et~al.}()\citenamefont {Danaila},
  \citenamefont {Anselmet}, \citenamefont {Zhou},\ and\ \citenamefont
  {Antonia}}]{Danaila1999}%
  \BibitemOpen
  \bibfield  {author} {\bibinfo {author} {\bibfnamefont {L.}~\bibnamefont
  {Danaila}}, \bibinfo {author} {\bibfnamefont {F.}~\bibnamefont {Anselmet}},
  \bibinfo {author} {\bibfnamefont {T.}~\bibnamefont {Zhou}},\ and\ \bibinfo
  {author} {\bibfnamefont {R.~A.}\ \bibnamefont {Antonia}},\ }\bibfield
  {title} {\bibinfo {title} {A generalization of {{Yaglom}}'s equation which
  accounts for the large-scale forcing in heated decaying turbulence},\ }\href
  {https://doi.org/10.1017/S0022112099005418} {\ \textbf {\bibinfo {volume}
  {391}},\ \bibinfo {pages} {359}}\BibitemShut {NoStop}%
\bibitem [{\citenamefont {Antonia}\ and\ \citenamefont
  {Burattini}(2006)}]{Antonia2006}%
  \BibitemOpen
  \bibfield  {author} {\bibinfo {author} {\bibfnamefont {R.~A.}\ \bibnamefont
  {Antonia}}\ and\ \bibinfo {author} {\bibfnamefont {P.}~\bibnamefont
  {Burattini}},\ }\bibfield  {title} {\bibinfo {title} {Approach to the 4/5 law
  in homogeneous isotropic turbulence},\ }\href
  {https://doi.org/10.1017/S0022112005008438} {\bibfield  {journal} {\bibinfo
  {journal} {Journal of Fluid Mechanics}\ }\textbf {\bibinfo {volume} {550}},\
  \bibinfo {pages} {175} (\bibinfo {year} {2006})}\BibitemShut {NoStop}%
\bibitem [{\citenamefont {Thiesset}\ \emph {et~al.}(2013)\citenamefont
  {Thiesset}, \citenamefont {Antonia}, \citenamefont {Danaila},\ and\
  \citenamefont {Djenidi}}]{Thiesset2013}%
  \BibitemOpen
  \bibfield  {author} {\bibinfo {author} {\bibfnamefont {F.}~\bibnamefont
  {Thiesset}}, \bibinfo {author} {\bibfnamefont {R.~A.}\ \bibnamefont
  {Antonia}}, \bibinfo {author} {\bibfnamefont {L.}~\bibnamefont {Danaila}},\
  and\ \bibinfo {author} {\bibfnamefont {L.}~\bibnamefont {Djenidi}},\
  }\bibfield  {title} {\bibinfo {title} {K\'arm\'an-{{Howarth}} closure
  equation on the basis of a universal eddy viscosity},\ }\href
  {https://doi.org/10.1103/PhysRevE.88.011003} {\bibfield  {journal} {\bibinfo
  {journal} {Physical Review E}\ }\textbf {\bibinfo {volume} {88}},\ \bibinfo
  {pages} {011003} (\bibinfo {year} {2013})}\BibitemShut {NoStop}%
\bibitem [{\citenamefont {Batchelor}(1951)}]{Batchelor1951}%
  \BibitemOpen
  \bibfield  {author} {\bibinfo {author} {\bibfnamefont {G.~K.}\ \bibnamefont
  {Batchelor}},\ }\bibfield  {title} {\bibinfo {title} {Pressure fluctuations
  in isotropic turbulence},\ }\href {https://doi.org/10.1017/S0305004100026712}
  {\bibfield  {journal} {\bibinfo  {journal} {Mathematical Proceedings of the
  Cambridge Philosophical Society}\ }\textbf {\bibinfo {volume} {47}},\
  \bibinfo {pages} {359} (\bibinfo {year} {1951})}\BibitemShut {NoStop}%
\bibitem [{\citenamefont {Lohse}\ and\ \citenamefont
  {{M{\"u}ller-Groeling}}(1995)}]{Lohse1995}%
  \BibitemOpen
  \bibfield  {author} {\bibinfo {author} {\bibfnamefont {D.}~\bibnamefont
  {Lohse}}\ and\ \bibinfo {author} {\bibfnamefont {A.}~\bibnamefont
  {{M{\"u}ller-Groeling}}},\ }\bibfield  {title} {\bibinfo {title} {Bottleneck
  {{Effects}} in {{Turbulence}}: {{Scaling Phenomena}} in r versus p
  {{Space}}},\ }\href {https://doi.org/10.1103/PhysRevLett.74.1747} {\bibfield
  {journal} {\bibinfo  {journal} {Physical Review Letters}\ }\textbf {\bibinfo
  {volume} {74}},\ \bibinfo {pages} {1747} (\bibinfo {year}
  {1995})}\BibitemShut {NoStop}%
\bibitem [{\citenamefont {Kurien}\ and\ \citenamefont
  {Sreenivasan}(2000{\natexlab{a}})}]{Kurien2000}%
  \BibitemOpen
  \bibfield  {author} {\bibinfo {author} {\bibfnamefont {S.}~\bibnamefont
  {Kurien}}\ and\ \bibinfo {author} {\bibfnamefont {K.~R.}\ \bibnamefont
  {Sreenivasan}},\ }\bibfield  {title} {\bibinfo {title} {Anisotropic scaling
  contributions to high-order structure functions in high-{{Reynolds}}-number
  turbulence},\ }\href {https://doi.org/10.1103/PhysRevE.62.2206} {\bibfield
  {journal} {\bibinfo  {journal} {Physical Review E}\ }\textbf {\bibinfo
  {volume} {62}},\ \bibinfo {pages} {2206} (\bibinfo {year}
  {2000}{\natexlab{a}})}\BibitemShut {NoStop}%
\bibitem [{\citenamefont {Dhruva}(2000)}]{Dhruva2000}%
  \BibitemOpen
  \bibfield  {author} {\bibinfo {author} {\bibfnamefont {B.~R.}\ \bibnamefont
  {Dhruva}},\ }\bibfield  {title} {\bibinfo {title} {An experimental study of
  high {{Reynolds}} number turbulence in the atmosphere},\ }\href@noop {}
  {\bibfield  {journal} {\bibinfo  {journal} {Ph.D. Thesis}\ ,\ \bibinfo
  {pages} {2717}} (\bibinfo {year} {2000})}\BibitemShut {NoStop}%
\bibitem [{\citenamefont {Meyers}\ and\ \citenamefont
  {Meneveau}(2008)}]{Meyers2008}%
  \BibitemOpen
  \bibfield  {author} {\bibinfo {author} {\bibfnamefont {J.}~\bibnamefont
  {Meyers}}\ and\ \bibinfo {author} {\bibfnamefont {C.}~\bibnamefont
  {Meneveau}},\ }\bibfield  {title} {\bibinfo {title} {A functional form for
  the energy spectrum parametrizing bottleneck and intermittency effects},\
  }\href {https://doi.org/10.1063/1.2936312} {\bibfield  {journal} {\bibinfo
  {journal} {Physics of Fluids}\ }\textbf {\bibinfo {volume} {20}},\ \bibinfo
  {pages} {065109} (\bibinfo {year} {2008})}\BibitemShut {NoStop}%
\bibitem [{\citenamefont {Antonia}(1982)}]{Antonia1982}%
  \BibitemOpen
  \bibfield  {author} {\bibinfo {author} {\bibfnamefont {R.~A.}\ \bibnamefont
  {Antonia}},\ }\bibfield  {title} {\bibinfo {title} {Reynolds number
  dependence of velocity structure functions in turbulent shear flows},\ }\href
  {https://doi.org/10.1063/1.863624} {\bibfield  {journal} {\bibinfo  {journal}
  {Physics of Fluids}\ }\textbf {\bibinfo {volume} {25}},\ \bibinfo {pages}
  {29} (\bibinfo {year} {1982})}\BibitemShut {NoStop}%
\bibitem [{\citenamefont {Tang}\ \emph {et~al.}(2019)\citenamefont {Tang},
  \citenamefont {Antonia}, \citenamefont {Djenidi},\ and\ \citenamefont
  {Zhou}}]{Tang2019}%
  \BibitemOpen
  \bibfield  {author} {\bibinfo {author} {\bibfnamefont {S.}~\bibnamefont
  {Tang}}, \bibinfo {author} {\bibfnamefont {R.~A.}\ \bibnamefont {Antonia}},
  \bibinfo {author} {\bibfnamefont {L.}~\bibnamefont {Djenidi}},\ and\ \bibinfo
  {author} {\bibfnamefont {Y.}~\bibnamefont {Zhou}},\ }\bibfield  {title}
  {\bibinfo {title} {Can small-scale turbulence approach a quasi-universal
  state?},\ }\href {https://doi.org/10.1103/PhysRevFluids.4.024607} {\bibfield
  {journal} {\bibinfo  {journal} {Physical Review Fluids}\ }\textbf {\bibinfo
  {volume} {4}},\ \bibinfo {pages} {024607} (\bibinfo {year}
  {2019})}\BibitemShut {NoStop}%
\bibitem [{\citenamefont {Vallikivi}\ \emph {et~al.}(2011)\citenamefont
  {Vallikivi}, \citenamefont {Hultmark}, \citenamefont {Bailey},\ and\
  \citenamefont {Smits}}]{Vallikivi2011}%
  \BibitemOpen
  \bibfield  {author} {\bibinfo {author} {\bibfnamefont {M.}~\bibnamefont
  {Vallikivi}}, \bibinfo {author} {\bibfnamefont {M.}~\bibnamefont {Hultmark}},
  \bibinfo {author} {\bibfnamefont {S.~C.~C.}\ \bibnamefont {Bailey}},\ and\
  \bibinfo {author} {\bibfnamefont {A.~J.}\ \bibnamefont {Smits}},\ }\bibfield
  {title} {\bibinfo {title} {Turbulence measurements in pipe flow using a
  nano-scale thermal anemometry probe},\ }\href
  {https://doi.org/10.1007/s00348-011-1165-4} {\bibfield  {journal} {\bibinfo
  {journal} {Experiments in Fluids}\ }\textbf {\bibinfo {volume} {51}},\
  \bibinfo {pages} {1521} (\bibinfo {year} {2011})}\BibitemShut {NoStop}%
\bibitem [{\citenamefont {Saffman}(1967)}]{Saffman1967}%
  \BibitemOpen
  \bibfield  {author} {\bibinfo {author} {\bibfnamefont {P.~G.}\ \bibnamefont
  {Saffman}},\ }\bibfield  {title} {\bibinfo {title} {The large-scale structure
  of homogeneous turbulence},\ }\href
  {https://doi.org/10.1017/S0022112067000552} {\bibfield  {journal} {\bibinfo
  {journal} {Journal of Fluid Mechanics}\ }\textbf {\bibinfo {volume} {27}},\
  \bibinfo {pages} {581} (\bibinfo {year} {1967})}\BibitemShut {NoStop}%
\bibitem [{\citenamefont {Sinhuber}\ \emph {et~al.}(2015)\citenamefont
  {Sinhuber}, \citenamefont {Bodenschatz},\ and\ \citenamefont
  {Bewley}}]{Sinhuber2015}%
  \BibitemOpen
  \bibfield  {author} {\bibinfo {author} {\bibfnamefont {M.}~\bibnamefont
  {Sinhuber}}, \bibinfo {author} {\bibfnamefont {E.}~\bibnamefont
  {Bodenschatz}},\ and\ \bibinfo {author} {\bibfnamefont {G.~P.}\ \bibnamefont
  {Bewley}},\ }\bibfield  {title} {\bibinfo {title} {Decay of {{Turbulence}} at
  {{High Reynolds Numbers}}},\ }\href
  {https://doi.org/10.1103/PhysRevLett.114.034501} {\bibfield  {journal}
  {\bibinfo  {journal} {Physical Review Letters}\ }\textbf {\bibinfo {volume}
  {114}},\ \bibinfo {pages} {034501} (\bibinfo {year} {2015})}\BibitemShut
  {NoStop}%
\bibitem [{\citenamefont {Benzi}\ \emph {et~al.}(1993)\citenamefont {Benzi},
  \citenamefont {Ciliberto}, \citenamefont {Tripiccione}, \citenamefont
  {Baudet}, \citenamefont {Massaioli},\ and\ \citenamefont
  {Succi}}]{Benzi1993}%
  \BibitemOpen
  \bibfield  {author} {\bibinfo {author} {\bibfnamefont {R.}~\bibnamefont
  {Benzi}}, \bibinfo {author} {\bibfnamefont {S.}~\bibnamefont {Ciliberto}},
  \bibinfo {author} {\bibfnamefont {R.}~\bibnamefont {Tripiccione}}, \bibinfo
  {author} {\bibfnamefont {C.}~\bibnamefont {Baudet}}, \bibinfo {author}
  {\bibfnamefont {F.}~\bibnamefont {Massaioli}},\ and\ \bibinfo {author}
  {\bibfnamefont {S.}~\bibnamefont {Succi}},\ }\bibfield  {title} {\bibinfo
  {title} {Extended self-similarity in turbulent flows},\ }\href
  {https://doi.org/10.1103/PhysRevE.48.R29} {\bibfield  {journal} {\bibinfo
  {journal} {Physical Review E}\ }\textbf {\bibinfo {volume} {48}},\ \bibinfo
  {pages} {R29} (\bibinfo {year} {1993})}\BibitemShut {NoStop}%
\bibitem [{\citenamefont {Kurien}\ and\ \citenamefont
  {Sreenivasan}(2000{\natexlab{b}})}]{kurienAnisotropicScalingContributions2000}%
  \BibitemOpen
  \bibfield  {author} {\bibinfo {author} {\bibfnamefont {S.}~\bibnamefont
  {Kurien}}\ and\ \bibinfo {author} {\bibfnamefont {K.~R.}\ \bibnamefont
  {Sreenivasan}},\ }\bibfield  {title} {\bibinfo {title} {Anisotropic scaling
  contributions to high-order structure functions in high-{{Reynolds}}-number
  turbulence},\ }\href {https://doi.org/10.1103/PhysRevE.62.2206} {\bibfield
  {journal} {\bibinfo  {journal} {Physical Review E}\ }\textbf {\bibinfo
  {volume} {62}},\ \bibinfo {pages} {2206} (\bibinfo {year}
  {2000}{\natexlab{b}})}\BibitemShut {NoStop}%
\bibitem [{\citenamefont {Biferale}\ and\ \citenamefont
  {Procaccia}(2005)}]{biferaleAnisotropyTurbulentFlows2005}%
  \BibitemOpen
  \bibfield  {author} {\bibinfo {author} {\bibfnamefont {L.}~\bibnamefont
  {Biferale}}\ and\ \bibinfo {author} {\bibfnamefont {I.}~\bibnamefont
  {Procaccia}},\ }\bibfield  {title} {\bibinfo {title} {Anisotropy in
  {{Turbulent Flows}} and in {{Turbulent Transport}}},\ }\href
  {https://doi.org/10.1016/j.physrep.2005.04.001} {\bibfield  {journal}
  {\bibinfo  {journal} {Physics Reports}\ }\textbf {\bibinfo {volume} {414}},\
  \bibinfo {pages} {43} (\bibinfo {year} {2005})},\ \Eprint
  {https://arxiv.org/abs/nlin/0404014} {arXiv:nlin/0404014} \BibitemShut
  {NoStop}%
\bibitem [{\citenamefont {Iyer}\ \emph {et~al.}(2020)\citenamefont {Iyer},
  \citenamefont {Sreenivasan},\ and\ \citenamefont {Yeung}}]{Iyer2020}%
  \BibitemOpen
  \bibfield  {author} {\bibinfo {author} {\bibfnamefont {K.~P.}\ \bibnamefont
  {Iyer}}, \bibinfo {author} {\bibfnamefont {K.~R.}\ \bibnamefont
  {Sreenivasan}},\ and\ \bibinfo {author} {\bibfnamefont {P.~K.}\ \bibnamefont
  {Yeung}},\ }\bibfield  {title} {\bibinfo {title} {Scaling exponents saturate
  in three-dimensional isotropic turbulence},\ }\href
  {https://doi.org/10.1103/PhysRevFluids.5.054605} {\bibfield  {journal}
  {\bibinfo  {journal} {Physical Review Fluids}\ }\textbf {\bibinfo {volume}
  {5}},\ \bibinfo {pages} {054605} (\bibinfo {year} {2020})}\BibitemShut
  {NoStop}%
\bibitem [{\citenamefont {Brasseur}\ and\ \citenamefont
  {Wei}(1994)}]{Brasseur1994}%
  \BibitemOpen
  \bibfield  {author} {\bibinfo {author} {\bibfnamefont {J.~G.}\ \bibnamefont
  {Brasseur}}\ and\ \bibinfo {author} {\bibfnamefont {C.}~\bibnamefont {Wei}},\
  }\bibfield  {title} {\bibinfo {title} {Interscale dynamics and local isotropy
  in high reynolds number turbulence within triadic interactions},\ }\href
  {https://doi.org/10.1063/1.868322} {\bibfield  {journal} {\bibinfo  {journal}
  {Physics of Fluids}\ }\textbf {\bibinfo {volume} {6}},\ \bibinfo {pages}
  {842} (\bibinfo {year} {1994})},\ \Eprint
  {https://arxiv.org/abs/https://doi.org/10.1063/1.868322}
  {https://doi.org/10.1063/1.868322} \BibitemShut {NoStop}%
\bibitem [{\citenamefont {Alexakis}\ \emph {et~al.}(2005)\citenamefont
  {Alexakis}, \citenamefont {Mininni},\ and\ \citenamefont
  {Pouquet}}]{Alexakis2005}%
  \BibitemOpen
  \bibfield  {author} {\bibinfo {author} {\bibfnamefont {A.}~\bibnamefont
  {Alexakis}}, \bibinfo {author} {\bibfnamefont {P.~D.}\ \bibnamefont
  {Mininni}},\ and\ \bibinfo {author} {\bibfnamefont {A.}~\bibnamefont
  {Pouquet}},\ }\bibfield  {title} {\bibinfo {title} {Imprint of large-scale
  flows on turbulence},\ }\href {https://doi.org/10.1103/PhysRevLett.95.264503}
  {\bibfield  {journal} {\bibinfo  {journal} {Phys. Rev. Lett.}\ }\textbf
  {\bibinfo {volume} {95}},\ \bibinfo {pages} {264503} (\bibinfo {year}
  {2005})}\BibitemShut {NoStop}%
\bibitem [{\citenamefont {Verma}\ and\ \citenamefont
  {Donzis}(2007)}]{Verma2007}%
  \BibitemOpen
  \bibfield  {author} {\bibinfo {author} {\bibfnamefont {M.~K.}\ \bibnamefont
  {Verma}}\ and\ \bibinfo {author} {\bibfnamefont {D.}~\bibnamefont {Donzis}},\
  }\bibfield  {title} {\bibinfo {title} {Energy {{Flux}} and {{Bottleneck
  Effect}} in {{Turbulence}}},\ }\bibfield  {journal} {\bibinfo  {journal}
  {arXiv:nlin/0510026}\ }\href {https://doi.org/10.1088/1751-8113/40/16/010}
  {10.1088/1751-8113/40/16/010} (\bibinfo {year} {2007}),\ \Eprint
  {https://arxiv.org/abs/nlin/0510026} {arXiv:nlin/0510026} \BibitemShut
  {NoStop}%
\bibitem [{\citenamefont {Leung}\ \emph {et~al.}(2012)\citenamefont {Leung},
  \citenamefont {Swaminathan},\ and\ \citenamefont {Davidson}}]{Leung2012}%
  \BibitemOpen
  \bibfield  {author} {\bibinfo {author} {\bibfnamefont {T.}~\bibnamefont
  {Leung}}, \bibinfo {author} {\bibfnamefont {N.}~\bibnamefont {Swaminathan}},\
  and\ \bibinfo {author} {\bibfnamefont {P.~A.}\ \bibnamefont {Davidson}},\
  }\bibfield  {title} {\bibinfo {title} {Geometry and interaction of structures
  in homogeneous isotropic turbulence},\ }\href
  {https://doi.org/10.1017/jfm.2012.373} {\bibfield  {journal} {\bibinfo
  {journal} {Journal of Fluid Mechanics}\ }\textbf {\bibinfo {volume} {710}},\
  \bibinfo {pages} {453} (\bibinfo {year} {2012})}\BibitemShut {NoStop}%
\bibitem [{\citenamefont {Bodenschatz}\ \emph
  {et~al.}(2014{\natexlab{b}})\citenamefont {Bodenschatz}, \citenamefont
  {Bewley}, \citenamefont {Nobach}, \citenamefont {Sinhuber},\ and\
  \citenamefont {Xu}}]{Bodenschatz2014a}%
  \BibitemOpen
  \bibfield  {author} {\bibinfo {author} {\bibfnamefont {E.}~\bibnamefont
  {Bodenschatz}}, \bibinfo {author} {\bibfnamefont {G.~P.}\ \bibnamefont
  {Bewley}}, \bibinfo {author} {\bibfnamefont {H.}~\bibnamefont {Nobach}},
  \bibinfo {author} {\bibfnamefont {M.}~\bibnamefont {Sinhuber}},\ and\
  \bibinfo {author} {\bibfnamefont {H.}~\bibnamefont {Xu}},\ }\bibfield
  {title} {\bibinfo {title} {Variable {{Density Turbulence Tunnel Facility}}},\
  }\href {https://doi.org/10.1063/1.4896138} {\bibfield  {journal} {\bibinfo
  {journal} {Review of Scientific Instruments}\ }\textbf {\bibinfo {volume}
  {85}},\ \bibinfo {pages} {093908} (\bibinfo {year} {2014}{\natexlab{b}})},\
  \Eprint {https://arxiv.org/abs/1401.4970} {arXiv:1401.4970} \BibitemShut
  {NoStop}%
\bibitem [{\citenamefont {Davidson}(2015)}]{Davidson2015}%
  \BibitemOpen
  \bibfield  {author} {\bibinfo {author} {\bibfnamefont {P.}~\bibnamefont
  {Davidson}},\ }\href@noop {} {\emph {\bibinfo {title} {Turbulence: {{An
  Introduction}} for {{Scientists}} and {{Engineers}}}}}\ (\bibinfo
  {publisher} {{Oxford University Press}},\ \bibinfo {year} {2015})\BibitemShut
  {NoStop}%
\bibitem [{\citenamefont {Kunkel}\ \emph {et~al.}(2006)\citenamefont {Kunkel},
  \citenamefont {Arnold},\ and\ \citenamefont {Smits}}]{Kunkel2006}%
  \BibitemOpen
  \bibfield  {author} {\bibinfo {author} {\bibfnamefont {G.}~\bibnamefont
  {Kunkel}}, \bibinfo {author} {\bibfnamefont {C.}~\bibnamefont {Arnold}},\
  and\ \bibinfo {author} {\bibfnamefont {A.}~\bibnamefont {Smits}},\ }\bibfield
   {title} {\bibinfo {title} {Development of {{NSTAP}}: {{Nanoscale Thermal
  Anemometry Probe}}},\ }in\ \href {https://doi.org/10.2514/6.2006-3718} {\emph
  {\bibinfo {booktitle} {36th {{AIAA Fluid Dynamics Conference}} and
  {{Exhibit}}}}}\ (\bibinfo  {publisher} {{American Institute of Aeronautics
  and Astronautics}},\ \bibinfo {address} {{San Francisco, California}},\
  \bibinfo {year} {2006})\BibitemShut {NoStop}%
\bibitem [{\citenamefont {Vallikivi}\ and\ \citenamefont
  {Smits}(2014)}]{Vallikivi2014}%
  \BibitemOpen
  \bibfield  {author} {\bibinfo {author} {\bibfnamefont {M.}~\bibnamefont
  {Vallikivi}}\ and\ \bibinfo {author} {\bibfnamefont {A.~J.}\ \bibnamefont
  {Smits}},\ }\bibfield  {title} {\bibinfo {title} {Fabrication and
  {{Characterization}} of a {{Novel Nanoscale Thermal Anemometry Probe}}},\
  }\href {https://doi.org/10.1109/JMEMS.2014.2299276} {\bibfield  {journal}
  {\bibinfo  {journal} {Journal of Microelectromechanical Systems}\ }\textbf
  {\bibinfo {volume} {23}},\ \bibinfo {pages} {899} (\bibinfo {year}
  {2014})}\BibitemShut {NoStop}%
\bibitem [{\citenamefont {Fan}\ \emph {et~al.}(2015)\citenamefont {Fan},
  \citenamefont {Arwatz}, \citenamefont {Van~Buren}, \citenamefont {Hoffman},\
  and\ \citenamefont {Hultmark}}]{Fan2015}%
  \BibitemOpen
  \bibfield  {author} {\bibinfo {author} {\bibfnamefont {Y.}~\bibnamefont
  {Fan}}, \bibinfo {author} {\bibfnamefont {G.}~\bibnamefont {Arwatz}},
  \bibinfo {author} {\bibfnamefont {T.~W.}\ \bibnamefont {Van~Buren}}, \bibinfo
  {author} {\bibfnamefont {D.~E.}\ \bibnamefont {Hoffman}},\ and\ \bibinfo
  {author} {\bibfnamefont {M.}~\bibnamefont {Hultmark}},\ }\bibfield  {title}
  {\bibinfo {title} {Nanoscale sensing devices for turbulence measurements},\
  }\href {https://doi.org/10.1007/s00348-015-2000-0} {\bibfield  {journal}
  {\bibinfo  {journal} {Experiments in Fluids}\ }\textbf {\bibinfo {volume}
  {56}},\ \bibinfo {pages} {138} (\bibinfo {year} {2015})}\BibitemShut
  {NoStop}%
\bibitem [{\citenamefont {Hutchins}\ \emph {et~al.}(2015)\citenamefont
  {Hutchins}, \citenamefont {Monty}, \citenamefont {Hultmark},\ and\
  \citenamefont {Smits}}]{Hutchins2015}%
  \BibitemOpen
  \bibfield  {author} {\bibinfo {author} {\bibfnamefont {N.}~\bibnamefont
  {Hutchins}}, \bibinfo {author} {\bibfnamefont {J.~P.}\ \bibnamefont {Monty}},
  \bibinfo {author} {\bibfnamefont {M.}~\bibnamefont {Hultmark}},\ and\
  \bibinfo {author} {\bibfnamefont {A.~J.}\ \bibnamefont {Smits}},\ }\bibfield
  {title} {\bibinfo {title} {A direct measure of the frequency response of
  hot-wire anemometers: Temporal resolution issues in wall-bounded
  turbulence},\ }\href {https://doi.org/10.1007/s00348-014-1856-8} {\bibfield
  {journal} {\bibinfo  {journal} {Experiments in Fluids}\ }\textbf {\bibinfo
  {volume} {56}},\ \bibinfo {pages} {18} (\bibinfo {year} {2015})}\BibitemShut
  {NoStop}%
\bibitem [{\citenamefont {Samie}\ \emph {et~al.}(2018)\citenamefont {Samie},
  \citenamefont {Hutchins},\ and\ \citenamefont {Marusic}}]{Samie2018a}%
  \BibitemOpen
  \bibfield  {author} {\bibinfo {author} {\bibfnamefont {M.}~\bibnamefont
  {Samie}}, \bibinfo {author} {\bibfnamefont {N.}~\bibnamefont {Hutchins}},\
  and\ \bibinfo {author} {\bibfnamefont {I.}~\bibnamefont {Marusic}},\
  }\bibfield  {title} {\bibinfo {title} {Revisiting end conduction effects in
  constant temperature hot-wire anemometry},\ }\href
  {https://doi.org/10.1007/s00348-018-2587-z} {\bibfield  {journal} {\bibinfo
  {journal} {Experiments in Fluids}\ }\textbf {\bibinfo {volume} {59}},\
  \bibinfo {pages} {133} (\bibinfo {year} {2018})}\BibitemShut {NoStop}%
\bibitem [{\citenamefont {Ashok}\ \emph {et~al.}(2012)\citenamefont {Ashok},
  \citenamefont {Bailey}, \citenamefont {Hultmark},\ and\ \citenamefont
  {Smits}}]{Ashok2012}%
  \BibitemOpen
  \bibfield  {author} {\bibinfo {author} {\bibfnamefont {A.}~\bibnamefont
  {Ashok}}, \bibinfo {author} {\bibfnamefont {S.~C.~C.}\ \bibnamefont
  {Bailey}}, \bibinfo {author} {\bibfnamefont {M.}~\bibnamefont {Hultmark}},\
  and\ \bibinfo {author} {\bibfnamefont {A.~J.}\ \bibnamefont {Smits}},\
  }\bibfield  {title} {\bibinfo {title} {Hot-wire spatial resolution effects in
  measurements of grid-generated turbulence},\ }\href
  {https://doi.org/10.1007/s00348-012-1382-5} {\bibfield  {journal} {\bibinfo
  {journal} {Experiments in Fluids}\ }\textbf {\bibinfo {volume} {53}},\
  \bibinfo {pages} {1713} (\bibinfo {year} {2012})}\BibitemShut {NoStop}%
\bibitem [{\citenamefont {Pao}(1965)}]{Pao1965}%
  \BibitemOpen
  \bibfield  {author} {\bibinfo {author} {\bibfnamefont {Y.-H.}\ \bibnamefont
  {Pao}},\ }\bibfield  {title} {\bibinfo {title} {Structure of {{Turbulent
  Velocity}} and {{Scalar Fields}} at {{Large Wavenumbers}}},\ }\href
  {https://doi.org/10.1063/1.1761356} {\bibfield  {journal} {\bibinfo
  {journal} {Physics of Fluids}\ }\textbf {\bibinfo {volume} {8}},\ \bibinfo
  {pages} {1063} (\bibinfo {year} {1965})}\BibitemShut {NoStop}%
\bibitem [{\citenamefont {Pope}\ and\ \citenamefont {Pope}(2000)}]{Pope2000}%
  \BibitemOpen
  \bibfield  {author} {\bibinfo {author} {\bibfnamefont {S.~B.}\ \bibnamefont
  {Pope}}\ and\ \bibinfo {author} {\bibfnamefont {S.~B.}\ \bibnamefont
  {Pope}},\ }\href@noop {} {\emph {\bibinfo {title} {Turbulent {{Flows}}}}}\
  (\bibinfo  {publisher} {{Cambridge University Press}},\ \bibinfo {year}
  {2000})\BibitemShut {NoStop}%
\bibitem [{\citenamefont {Monin}\ and\ \citenamefont
  {Yaglom}(1975)}]{Monin1975}%
  \BibitemOpen
  \bibfield  {author} {\bibinfo {author} {\bibfnamefont {A.~S.}\ \bibnamefont
  {Monin}}\ and\ \bibinfo {author} {\bibfnamefont {A.}~\bibnamefont {Yaglom}},\
  }\href@noop {} {\emph {\bibinfo {title} {Statistical Fluid Mechanics:
  Mechanics of Turbulence. 2 2}}},\ edited by\ \bibinfo {editor} {\bibfnamefont
  {J.~L.}\ \bibnamefont {Lumley}}\ (\bibinfo  {publisher} {{MIT Pr.}},\
  \bibinfo {address} {{Cambridge, Mass.}},\ \bibinfo {year} {1975})\BibitemShut
  {NoStop}%
\end{thebibliography}%


%apsrev4-2.bst 2019-01-14 (MD) hand-edited version of apsrev4-1.bst
%Control: key (0)
%Control: author (8) initials jnrlst
%Control: editor formatted (1) identically to author
%Control: production of article title (0) allowed
%Control: page (0) single
%Control: year (1) truncated
%Control: production of eprint (0) enabled
%


%apsrev4-2.bst 2019-01-14 (MD) hand-edited version of apsrev4-1.bst
%Control: key (0)
%Control: author (8) initials jnrlst
%Control: editor formatted (1) identically to author
%Control: production of article title (0) allowed
%Control: page (0) single
%Control: year (1) truncated
%Control: production of eprint (0) enabled
%
%\begin{figure*}
%  \includegraphics[width=\textwidth]{Constants.pdf}
%  \caption{(A) $S_2(r)$ compensated by the dimensional prediction for %the same $R_\lambda$ as in Fig. \ref{fig:Fig1} (A). Neither a %universal form nor the anticipated inertial range plateau can be %identified. (B) $C_2$ for different $R_\lambda$. The data is %scattered, but shows a clear $R_\lambda$-dependence. (C) $S_3(r)$ %compensated by the dimensional prediction for the same $R_\lambda$ %as in Fig. \ref{fig:Fig1} (A). Neither a universal form nor the %anticipated inertial range plateau can be identified, but the peaks %begin to collapse at the highest Reynolds numbers. This is %confirmed by (D), where we illustrate the approach of $S_3(r)$ %towards the $4/5$-law with increasing $R_\lambda$. 
%  Black lines correspond to the predictions by the physical model of %\citet{Yang2018}. In (A) and (C) the model prediction for %$R_\lambda=5175$ are shown.  
%  Closed circles were corrected for finite probe resolution, temporal %filtering, and a fluctuation advection velocity, no corrections %were applied to the open circles. Errorbars denote statistical %errors in $S_3$ and a 4\% error (8\% on uncorrected data) on the %dissipation rate $\varepsilon$ and $r$. }
%  \label{fig:Constants}
%\end{figure*}

\appendix
%%%%%%%%%%%%
%%%%METHODS/SI
%%%%%%%%%%%%%
\section{Methods}
\subsection{The Max Planck Variable Density Turbulence Tunnel}
The Variable Density Turbulence Tunnel (VDTT) \cite{Bodenschatz2014a} is a closed-loop wind tunnel, which can be operated with any non-corrosive gas at pressures up to 15 bar. For the experiments presented here it was operated with sulphur-hexaflouride (SF\textsubscript{6}), which offers a low kinematic viscosity that decreases with density while being relatively harmless and inert.
The Reynolds number of the flow in the VDTT can be finely adjusted in three largely independent ways up to levels typical for atmospheric turbulence: (i) the large-scale forcing with a novel active grid, (ii) the mean flow speed $U$ up to 5.5\,m/s by adjusting the rotation frequency of its fan, and (iii) the kinematic viscosity $\nu$ by changing the static pressure. 

Flow structures of variable size are introduced using a mosaic-like arrangement of individually controllable paddles ("active grid"). It allows us to obstruct the flow on finely adjustable time- and length scales \cite{Griffin2019,Kuchler2019}. The resulting grid length scale is indicated in Fig. \ref{fig:ExperimentalConditions} as red vertical lines. In this way we control the energy injection scale between about $0.1 \mathrm{m} \lesssim L \lesssim 0.6 \mathrm{m}$. $L$ is indicated as short black vertical lines in Fig.~\ref{fig:ExperimentalConditions}.

The small kinematic viscosity of pressurized SF$_6$ permits the existence of very small flow structures. The size of these structures scales with the viscous length scale $\eta=(\nu^3/\varepsilon)^{1/4}$, where $\varepsilon = 15\nu \langle (\partial u/\partial x)^2\rangle$. For the range of ambient pressures 1 bar $< p <$ 15 bar, this viscous length is between $250 \mathrm{\mu m} \gtrsim \eta \gtrsim 10 \mathrm{\mu m}$.

\begin{figure}
    \includegraphics[width=\columnwidth]{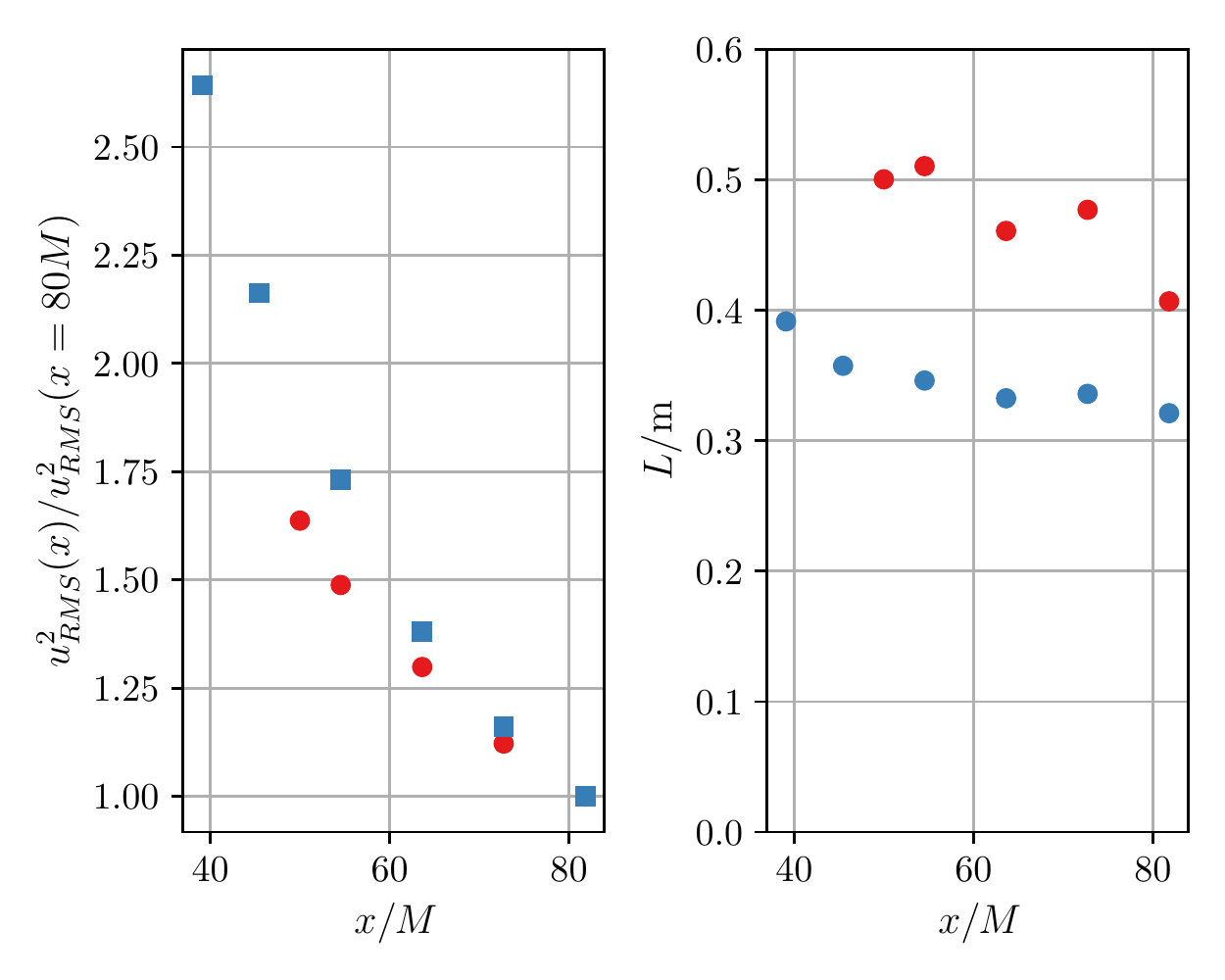}
    \caption{
    %\centering
    Development of the turbulent kinetic energy (left) and integral length scale $L$ (right) for different distances from the active grid. The distances are normalised by the active grid paddle dimensions. $L$ was estimated using $\int_0^{r_0} \! C_{11} (r)/u_{RMS}^2 \, dr$ with $r_0$ the first zero-crossing of the correlation function.}    
    \label{fig:DecayPlot}
\end{figure}

In our experiment, the turbulent kinetic energy $u_{RMS}^2$ decays along the length of the measurement section, but the integral length scale $L$ remains constant or also decays over time (see Fig.~\ref{fig:DecayPlot}). This is in contrast to freely decaying turbulence, where $L$ grows with time \cite{Davidson2015,Sinhuber2015}. We believe that the boundaries of the measurement section with cross-section 1.2\,m $\times$ 1.5\,m (with 0.1\,m $\lesssim L \lesssim $ 0.6\,m) suppresses this growth. 
We found this to be relatively independent of the way we estimate $L$. We chose to use $L=\int_0^{r_s} \langle u(x)u(x+r) \rangle/u_{RMS}^2 dr$ with $\langle u(x)u(x+r_s) \rangle=0$. Other definitions of $L$ impact the results at small $R_\lambda$ and the scatter of the data otherwise.

\subsection{Measurement Technology and Data Analysis}
\begin{figure}
    \centering
    \includegraphics[width=\columnwidth]{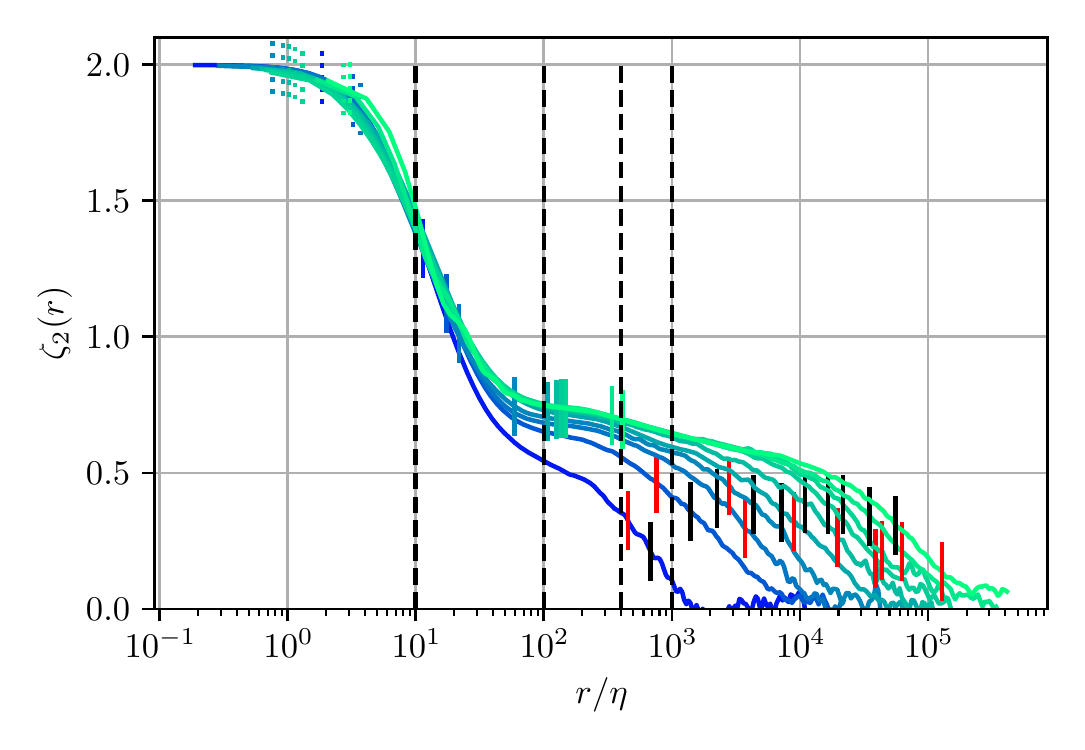}
    \caption{Same as Fig.~\ref{LocalSlopeCollapseEta} (A) with the addition of probe length (dotted vertical lines), the value of $r/\eta$ corresponding to a measurement frequency of 1\,kHz through Taylor's Hypothesis (vertical lines), the values of $r_0/\eta$ chosen to assemble Fig.~\ref{LocalSlopeCollapseEta} (D) (dashed black lines), the length of the energy injection scale (vertical black lines), and the grid length scale (red lines). }
    \label{fig:ExperimentalConditions}
\end{figure}

We record time series of hot-wire signals and convert them into one-dimensional flow fields assuming that the turbulent fluctuations are passively advected across the sensor by the mean flow $U$. Thus, a time step $\Delta t$ is converted to a spatial increment $\Delta x = U \Delta t$ \cite{Taylor1935}. We use a commercial constant temperature anemometer (Dantec StreamWare) to drive and acquire data from Nanoscale Thermal Anemometry Probes (NSTAP) provided by Princeton University \cite{Kunkel2006,Vallikivi2014,Fan2015}. These ultra-small hot wire probes average the flow field over a length of only 30\,$\mu$m, which is sufficient for this experiment. For flows where the viscous length scales are larger, we also use commercial hot wires from Dantec Dynamics with sensing length 450\,$\mu$m ($\gtrsim 4 \eta$). The probe length is indicated by dotted vertical lines in Fig. \ref{fig:ExperimentalConditions} and far away from the region of interest.

To achieve converged statistics the data was acquired for $10^3$ - $10^4$
eddy turnover times (up to 8 hours)  between $150 < R_\lambda < 6000$. 

The frequencies (and wavenumbers) encountered in the measurements presented here are generally in a range that is not particularly demanding for this combination of sensor and anemometer circuitry \cite{Hutchins2015,Samie2018a,Ashok2012}. The temporal resolution is determined by the noise filtering frequency and the frequency response of the measurement system. The frequency response of the system is not perfectly flat anymore starting around 1\,kHz \cite{Hutchins2015}. The range of scales we are interested in is therefore in the flat part of the frequency response curve. To illustrate this, the length scales corresponding to a measurement frequency of 1 kHz are indicated in Fig.~\ref{fig:ExperimentalConditions} as vertical lines in the color of the corresponding $\zeta_2(r)$. The noise filtering frequency is always at frequencies above 1kHz. 

The experiments presented here were taken under different ambient pressures and different active grid forcing schemes to allow for a careful check of the hot wire fidelity. We thus ensure the robustness of the results against probe- or flow geometry-induced biases. We emphasise that all conclusions presented here are independent of the frequencies where turbulent fluctuations are measured, the dissipation length scale, and the active grid forcing. 

\subsection{Fits to the Model Spectrum \cite{Yang2018}}\label{app:ModelSpec}
The evolution equation of the velocity energy spectrum $E(k,t)$ can be derived directly from the Navier-Stokes-Equation in the isotropic case and is known as the Karman-Howarth-Lin equation.
\begin{equation}
  \partial_t E(k,t) = -\partial_k\Pi(k,t)-2\nu k^2E(k,t).
  \label{KarmanHowarth}
\end{equation}
The first term on the RHS describes the nonlinear transfer of energy from small to large wavenumbers and ultimately prevents the closure of the equation, since it is a third-order term. The Pao closure \cite{Pao1965} used in the model by Yang et al. \cite{Yang2018} assumes that the transfer term $\Pi$ is local in wavenumber space and has a self-similar form:
\begin{equation}
    \Pi(k,t)=C_0\varepsilon^{1/3}k^{5/3}E(k,t)
\end{equation}
The second term on the RHS of (\ref{KarmanHowarth}) represents the viscous dissipation at the smallest flow scales. This yields a closed form of the Karman-Howarth-Lin equation. The model by Yang et al. further assumes that the energy spectrum can be assembled by a large scale term $f_L(kL)$, a small scale term $f_\eta(k\eta)$, and a self-similar inertial range:
\begin{equation}
    E(k,t)= C_k \varepsilon^{2/3} k^{5/3} f_\eta(k\eta) f_L(k L)
    \label{eq:PaoHyp}
\end{equation}
These assumptions are now combined with a general, self-similar decay of turbulent kinetic energy. In the case of a confined domain, where the parameter describing $d L/d t$ tends to zero, this model predicts the energy spectrum
\begin{equation}
  E(k) \sim \frac{-A_K}{C}(kL)^{-(\zeta_{2F}+1)}e^{(3A_K/2C)(kL)^{-2/3}}e^{-(1.5/C)(k\eta)^{4/3}}.
  \label{eqn:YangModel}
\end{equation}

For the purpose of measuring a scaling exponent, we replaced the term $(kL)^{-5/3}$ used in the original formulation of the spectrum with $(kL)^{-(\zeta_{2F}+1)}$, where the fitting parameter $\zeta_{2F}$ is the inertial range scaling exponent for the second order structure function\cite{Pope2000}. The parameters $C$ and $A_K$ are related through $C=-A_K(6/\pi)^{1/3}$. In practice, $A_K$ describes the large-scale part of the energy spectrum, which is heavily influenced by the decay.
 
The one-dimensional versions of $S_2$ and $E(k)$ are related through the following integral transform \cite{Monin1975}:
\begin{equation}
 S_2(r)=\int_0^{\infty} \! E(k) \left(\frac{1}{3}+\frac{\cos(kr)}{(kr)^2}-\frac{\sin(kr)}{(kr)^3}\right) dk.
  \label{eqn:IntegralTransform}
\end{equation}
To obtain the fits shown in Fig. \ref{fig:Fits}, we have searched for parameters $A_K$, and $\zeta_{2F}$ that yield best fits of the logarithmic derivative of eq. (\ref{eqn:IntegralTransform}) to the experimentally measured $\zeta_2(r)$.

It can be shown that $C=-A_K(6/\pi)^{1/3}$. This quantity is related to the dissipation constant $C_{\varepsilon}=\varepsilon L/u^3$ relating the large scale energy injection and the small scale energy transfer rate $\varepsilon$. $A_K$ is the non-dimensionalized time-evolution of the energy spectrum prefactor $d (C_K \varepsilon^{2/3})/dt$, which is a free parameter.

The energy transfer spectrum $\Pi(k)$ is related to $S_3$ via
\begin{equation}
    S_3 = 12\int_0^{\infty} \! \frac{1}{k^2} \frac{\partial \Pi}{\partial k} \frac{d}{dr}\left(\frac{1}{3}+\frac{\cos(kr)}{(kr)^2}-\frac{\sin(kr)}{(kr)^3}\right) dk. 
\end{equation}

The second derivative has been estimated by a Taylor expansion for $kr<0.001$. Therefore, the model (\ref{eqn:YangModel}) in combination with its underlying closure hypothesis eq.~(\ref{eq:PaoHyp}) implicitly predicts $S_3$. 
Note that strictly speaking the combination of the intermittency-corrected model eq.~(\ref{eqn:YangModel}) and the K41-type closure eq.~(\ref{eq:PaoHyp}) yields a third order exponent $\zeta_3$ slightly different from 1. It is reassuring to see that instead leaving the 5/3-term in eq.~(\ref{eq:PaoHyp}) as a generic scaling and fitting the resulting model to $S_3$ yields $\Pi \approx$ const. in the inertial range so that $S_3 \sim r$. 

\subsection{Dependence of $F_2$ on $\mu_2$}
\begin{figure*}
    \centering
    \includegraphics[width=\textwidth]{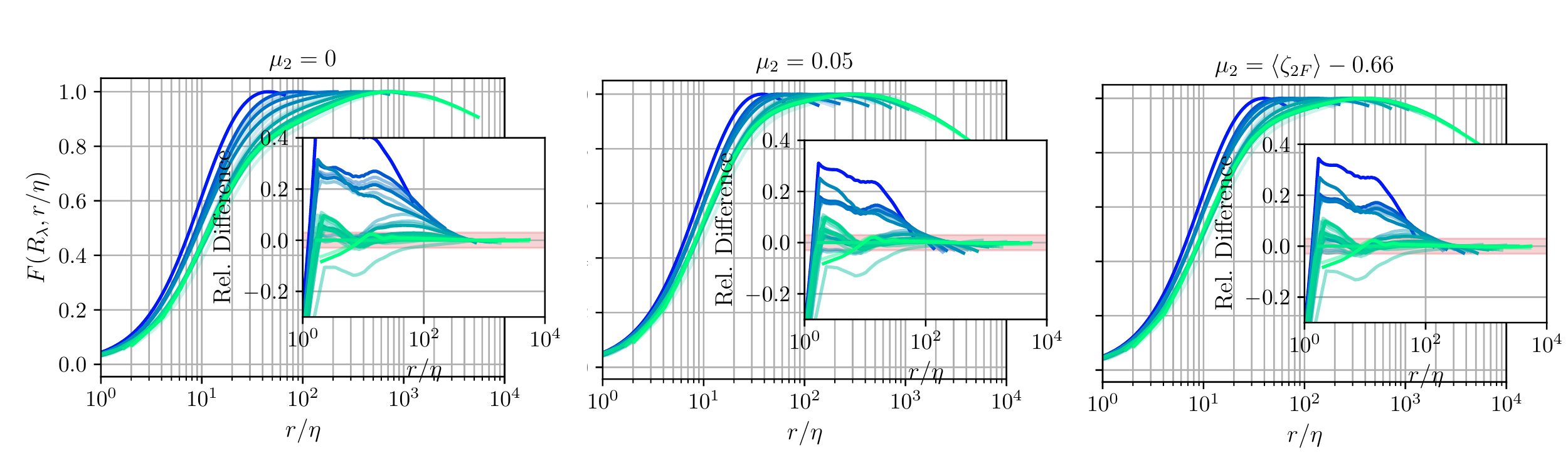}
    \caption{$F_2$ for different values of $\mu_2$. The K41 estimate $0.66$ provides the least degree of universality. No substantial differences appear in the most likely range of $\mu_2$ between 0.69 and 0.71.}
    \label{fig:MuStudy}
\end{figure*}
The precise form of $F_2(R_\lambda,r/\eta)$ depends on the choice of $\mu_2$, which is subject to systematic and statistical measurement errors. A poor estimate of $\mu_2$ might thus distort $F_2$ and disguise departures from universality of this function. We have recomputed the lower plot of Fig. \ref{fig:FitResults} for different values of $\mu_2$ and present the results in Fig. \ref{fig:MuStudy}. $F_2$ does not vary noticably for exemplary values of $\mu_2$ within the most likely true range. It becomes clear that the K41 prediction $\mu=0$ does not produce a universal form of $F_2$, which is expected in the light of the vast evidence in favour of intermittency corrections. Values of $\mu$ within its likely range between 0.68 and 0.71 yield very similar values of $F_2$. Most importantly its universality is not impacted significantly.

\end{document}